\PassOptionsToPackage{colorlinks=true,allcolors=black}{hyperref}
\documentclass[journal]{new-aiaa}

\usepackage[utf8]{inputenc}
\usepackage{bm}

\usepackage{mathtools}
\usepackage{amsfonts} % for \mathbb
\usepackage{comment}

\usepackage{graphicx}
\usepackage{booktabs}
\usepackage{longtable,tabularx}
\usepackage{multirow}
\usepackage{subcaption}

\usepackage[version=4]{mhchem}
\usepackage{siunitx}
\usepackage{algorithm}
\usepackage{algorithmic}

\usepackage{textcomp}
\usepackage{xcolor}

\usepackage[numbers,sort&compress]{natbib}

\usepackage{hyperref}
\hypersetup{colorlinks=true,allcolors=black}

\newcommand{\bmat}[1]{\begin{bmatrix}#1\end{bmatrix}}

\newcommand{\mbf}{\mathbf}

\newcommand{\tr}{^{\mathrm{T}}}

\title{Polynomial Updates for the Unscented Kalman Filter}

\author{Chiran Binnu Cherian\footnote{Graduate Student, Department of Aerospace Engineering, cbckbc@iastate.edu, and Student Member AIAA.} and Simone Servadio\footnote{Assistant Professor, Department of Aerospace Engineering, servadio@iastate.edu, and AAS/AIAA Member}}
\affil{Iowa State University, Ames, IA, 50011-2030}

\begin{document}

\maketitle

\begin{abstract}
Most nonlinear filters used in spacecraft navigation are based on a linear approximation of the optimal minimum mean square error estimator. The Unscented Kalman Filter (UKF) handles nonlinear dynamics through a sigma-point transform, but the resulting state estimate remains a linear function of the measurement. This paper proposes a polynomial approximation of the optimal Bayesian update, leading to a Polynomial Unscented Kalman Filter that retains the structure of the standard UKF but enriches the measurement update with higher-order (polynomial) terms. To compute the moments required by this polynomial estimator, we employ a Conjugate Unscented Transformation (CUT), which accurately captures higher-order central moments of the state and measurement. Numerical examples, including Clohessy-Wiltshire and Circular Restricted 3-Body dynamics with non-Gaussian measurement noise, illustrate that the resulting polynomial-CUT filters improve both state estimation accuracy and covariance consistency when compared with their linear counterparts.
\end{abstract}

\section{Introduction}
\lettrine{T}{he} Kalman filter provides the canonical solution for linear Gaussian systems, yielding an exact, recursive computation of the posterior mean and covariance \cite{kalman1960new}. Its success led to widespread adoption in guidance, navigation, and control and motivated systematic treatments of estimation under uncertainty \cite{gelb1974applied}. Most flight-relevant models, however, are nonlinear and often operate outside locally Gaussian regimes, which compromises the optimality guaranties of the linear filter.

To address nonlinearity, the Extended Kalman Filter (EKF) linearizes the dynamics and measurement models about the current estimate and applies the Kalman machinery to this first-order approximation \cite{gelb1974applied}. Although simple and efficient, the EKF can suffer from truncation bias and covariance inconsistency under strong nonlinearity, as emphasized in orbit and attitude applications \cite{junkins2004nonlinear}. The Unscented Kalman Filter (UKF) replaces Jacobian linearization with deterministic sigma-point quadrature that more accurately transports mean and covariance through nonlinear maps while retaining Gaussian output assumptions \cite{julier2004unscented}. Related sigma-point and ensemble variants; Cubature KF (CKF), Ensemble KF (EnKF), and Central-Difference KF (CDKF) offer alternative deterministic or ensemble sampling strategies that target third-degree accuracy or finite-difference approximations \cite{arasaratnam2009cubature, servadio2021differential, valli2012gaussian, schei1997finite}. Collectively, these approaches improve propagation fidelity but still perform an affine measurement update, which can under-fit curvature in the state–measurement relation.
Beyond first-order or sigma-point propagation, higher-order tools have been developed to move uncertainty through nonlinear dynamics more faithfully. State Transition Tensors (STT) capture higher-order sensitivities on the flow map, enabling richer moment transport for dynamic systems \cite{majji2008high}. Differential Algebra (DA) provides algebraic operators that propagate polynomial representations of uncertainty, delivering accurate nonlinear mappings for celestial mechanics and related problems \cite{valli2013nonlinear}. These methods primarily enhance the time-update (prediction) step; in many implementations, the measurement update remains affine.

When the posterior departs strongly from Gaussianity (for example, heavy tails, skewness, or multimodality), approximation families that go beyond single-Gaussian representations become attractive. Gaussian Sum Filters (GSF) model the state density as a mixture of Gaussians and update each component recursively, approximating nonlinear posteriors with controllable expressiveness \cite{sorenson1971recursive, alspach2003nonlinear}. Particle Filters (PF), including bootstrap and Gaussianized variants, perform Bayesian recursion via weighted particles, accommodating arbitrary likelihoods and dynamics \cite{gordon1993novel, hutter2003gaussian}. Recent work has focused on proposal design and likelihood handling for challenging aerospace problems \cite{servadio2024likelihood}. Although these mixture/particle approaches can approximate the nonlinear conditional mean with high fidelity, they typically incur higher computational cost and face degeneracy or scaling challenges in higher dimensions.

Kalman-type filters built on linear or sigma-point propagation ultimately apply an affine correction at the measurement step. This is efficient and convenient, but it leaves performance on the table whenever the state–measurement relation exhibits significant curvature or the prior/likelihood departs from local Gaussianity. The symptoms are familiar to practitioners: biased posterior means, mis-calibrated covariances, and residual coverage that drifts from nominal bounds in strongly nonlinear regimes.

The quadratic update replaces the affine correction with a second-order approximation to the optimal Bayesian estimator. Intuitively, instead of mapping innovations to state corrections along a single linear direction, the update bends with the local geometry of the joint state-measurement distribution. This preserves the recursive structure, state covariance bookkeeping, and low-latency character of Kalman filtering while capturing curvature that affine updates cannot. Earlier polynomial-update formulations demonstrated this promise, but often required lifted measurements, struggled with vector observations, or demanded explicit transport of high-order moments \cite{de1995optimal, carravetta1997polynomial, germani2005polynomial}. Subsequent work addressed these practical hurdles using Taylor-based updates and efficient moment handling, avoiding complicated  systems while retaining the recursive filter form \cite{servadio2020recursive, servadio2020nonlinear, isserlis1918formula, servadio2021estimation}.

We instantiate this idea in two forms. The Quadratic Unscented Kalman Filter (QUKF) constructs the second-order correction without analytic derivatives, using sigma-point statistics near the predicted measurement to identify the necessary curvature terms. Operationally, QUKF preserves the UKF workflow by predicting via sigma points, assembling the required statistics, and applying a closed-form quadratic correction so that it drops into existing unscented pipelines with minimal refactoring.

Recent developments have consolidated these ideas into unified formulations that present both quadratic extended and quadratic unscented updates within a common framework. In particular, we provided a treatment of Quadratic Extended and Unscented Kalman Filter updates that systematize the derivation, clarifies the required statistics, and outlines implementation pathways that fit naturally into EKF and UKF-style pipelines \cite{servadio2025quadratic_article}. This line of work emphasizes practical considerations such as avoiding explicit high-order moment storage, handling vector measurements without singular lifts, and preserving square-root numerics for stability.

A crucial design choice is augmentation. Practical estimation problems demand that process and measurement noise, and often slowly varying biases or constant parameters, be transported through the same nonlinear maps as the state. We therefore adopt an augmented formulation in which the quadratic update is applied after jointly propagating state and noise variables. This ensures that the captured curvature reflects not only nominal dynamics and sensor models but also how uncertainties themselves flow through those models.

Many guidance, navigation, and control problems require propagating uncertainty through nonlinear models, which reduces to evaluating multivariate expectation integrals that rarely admit closed-form solutions outside linear–Gaussian cases. In practice, these integrals dominate the computational footprint of uncertainty quantification and filtering methods such as stochastic collocation, polynomial chaos, and Gaussian mix approaches \cite{madankan2014hazard, xiu2002wiener, madankan2013pcbayes, terejanu2008gmm, terejanu2011agsf, vishwajeet2014nonlinear}. When analytic expressions are not available, numerical quadrature is required. A wide spectrum of techniques exists for Gaussian and uniform priors such as Monte Carlo sampling, Gauss–Hermite and Gauss–Legendre rules, sparse-grid constructions, and sigma-point or cubature methods such as unscented and cubature transforms \cite{arasaratnam2009cubature, stroud1971multiple, stroud1966gaussian, gerstner1998sparse, julier2000new, wu2006numerical, stroud1960degree2}. All approximate the expectation by a weighted sum of function evaluations and they differ mainly in how they place points and assign weights.

Classical Gaussian quadrature achieves high accuracy, but its cost grows sharply with dimension and often introduces negative weights, which can damage stability and inflate the truncation error for nonpolynomial integrands \cite{stroud1966gaussian, gerstner1998sparse, mcnamee1967fully, wu2006numerical}. Sparse grids alleviate growth, but still suffer from anisotropy and complexity in the rule-design in high dimensions \cite{gerstner1998sparse}. Sigma-point rules strike a pragmatic balance: the unscented transform provides third-degree exactness with a compact, symmetric set of points that also cancel all odd moments, and reduced or simplex variants shrink the set further at the expense of those higher-order symmetries \cite{julier2000new, julier2002reduced}. The Kalman cubature filter adopts an alternative, equally third-degree arrangement with a similar point count \cite{arasaratnam2009cubature}. These well-known rules can be viewed as special cases of more general symmetric cubature structures developed for multivariate integration \cite{stroud1960degree2}, and have proven attractive for real-time nonlinear filtering.

The Conjugate Unscented Transformation (CUT) extends this family by constructing sets of sigma-points of higher-degree, strictly symmetric whose size grows only polynomially with dimension \cite{adurthi2015cut, adurthi2012cut, adurthi2012ms, adurthi2018cut_applications}. The key idea is to assemble conjugate tuples of axis-aligned patterns with balanced signs so that all odd moments vanish by symmetry while targeted even moments (up to fourth, sixth, or eighth order, depending on the chosen family) are matched exactly. As a result, CUT increases moment fidelity beyond the third degree without resorting to exponentially large grids. It retains the familiar workflow of unscented-style methods: compute a square-root factor of the covariance, generate a canonical point set, push points through the nonlinear maps, and form weighted reductions. Because the defining radii and weights are fixed once per degree and reused at every step, the implementation effort is modest. In addition, CUT’s symmetry typically yields nonnegative weights, which aids numerical stability and helps maintain positive-definite covariance updates \cite{adurthi2012cut, adurthi2012ms, mcnamee1967fully, wu2006numerical}.

CUT has been demonstrated in estimation and control tasks and in demanding space applications, notably accurate conjunction analysis under realistic uncertainties \cite{adurthi2015cut, adurthi2018cut_applications}. Variants tailored to non-Gaussian priors, including uniform distributions, further extend its applicability beyond the Gaussian assumption commonly made in sigma-point filters \cite{adurthi2013cut_uniform}. In this work, we employ an arbritary high-order CUT within an augmented state formulation so that the physical state, process noise, and measurement noise are transported through the same nonlinear maps. This choice offers a favorable accuracy–cost trade since it tightens predicted means, covariances, and cross-covariances in regimes where third-degree rules under-represent curvature, yet preserves the runtime and code simplicity valued in on-board implementations \cite{adurthi2015cut, adurthi2012cut, adurthi2013cut_uniform, adurthi2012ms, adurthi2018cut_applications}.

The quadratic update, or its polynomial generalization, and CUT are complementary. CUT improves the fidelity of predicted moments by enforcing higher-order symmetry conditions with polynomial growth in point count. The quadratic update then converts those more accurate prediction and measurement moments into a curvature-aware correction. Empirically, this pairing, QUKF with CUT in an augmented setting, produces more faithful tracking of the spread of the joint distribution, improved covariance calibration, and lower RMSE, as this paper will show in numerical applications across scalar toy problems, nonlinear Clohessy–Wiltshire dynamics with non-Gaussian noise, and Circular Restricted Three-Body dynamics.

The scope and limitations are clear. The quadratic update remains a single-mode approximation. In the presence of severe multimodality or extreme heavy tails, mixture-based or particle methods may be preferable. However, within the broad class of aerospace estimation problems characterized by pronounced nonlinearity and moderate non-Gaussianity, a quadratic update coupled with higher-order sigma-point propagation offers a practical accuracy–cost sweet spot that preserves recursive simplicity. Regardless, the proposed methodology can be generalized with multiple model mathematics without any loss of generality .

%%%%%%%%%%%%%%%%%%%%%%%%%%%%%%%%%%%%%%%%%%%%%%%%%%%%%%%%%%%%%%%%
\section{The Optimal Polynomial Estimator} \label{sec1}
We interpret the polynomial estimators considered in this work as approximations of the minimum mean square error (MMSE) estimator within a finite-dimensional family of polynomial functions of the measurement. In general, we restrict the estimate to have the form 
\[ 
    g_N(\mbf y) = \mathbb{E}[\mbf x] + \mbf K_N \,\psi_N(\delta \mbf y),
\] 
where $\delta \mbf y = \mbf y - \mathbb{E}[\mbf y]$ denotes the measurement deviation and $\psi_N(\delta \mbf y)$ collects all centralized monomials in $\delta \mbf y$ up to order $N$ (only linear terms for $N=1$, linear and quadratic terms for $N=2$, etc.). Among all such degree-$N$ polynomials, the optimal estimator $g_N^*$ is the one that minimizes the mean square error, i.e., the $L^2$ projection of $\mbf x$ onto $\mathrm{span}\{\psi_N(\delta \mbf y)\}$. By the orthogonality principle, this projection is characterized by 
\[ 
    \mathbb{E}\bigl[(\mbf x - g_N^*(\mbf y))\,\psi_N(\delta \mbf y)^T\bigr] = \mbf 0, 
\] 
which leads to a set of normal equations for the gain matrix $\mbf K_N$ in terms of joint moments of $\mbf x$ and the measurement $\mbf y$. The classical linear estimator corresponds to $N=1$, yielding $g_L(\mbf y)$ in Eq. \eqref{eq:linest}, while the quadratic case $N=2$ produces the quadratic family $g_Q(\mbf y)$ in terms of $\delta \mbf y$ and its Kronecker square $\delta \mbf y^{[2]}$, as derived in the next subsection.

\subsection{The Optimal Quadratic Estimator}
Previous works \cite{servadio2020recursive, servadio2020nonlinear, servadio2021estimation} derived a new series of polynomial estimators in which the state estimate is a polynomial function of the given measurement. They make use of Differential Algebra techniques to calculate high-order central moments. In this work, we obtain the quadratic estimator from the definition of the orthogonality principle, but apply the unscented transformation to obtain the high-order gains and terms, so that the final results resemble the structure of the Unscented Kalman Filter, but with the quadratic term. 

Minimum Mean Square Error (MMSE) estimators aim at minimizing the expected value of the error square; thus, their formulation leads to unbiased filters where the mean of the error square coincides with the error covariance. Whenever a linear function in the measurement is used, as a family of estimators, we obtain a linear representation of the MMSE, the LMMSE. These estimators have the form of
\begin{equation}
  g_L(\mbf y) = A + \mathbb{E}[\mbf x] + B\,(\mbf y - \mathbb{E}[\mbf y])  \label{eq:linest}
\end{equation}
and are very common. Equation \eqref{eq:linest} is used by the UKF and other linear estimators in their state update step. Their estimate changes according to the technique used to evaluate the constants $A$ and $B$, dealing with the nonlinearities of the measurement model. 

This work applies a quadratic approximation to the MMSE, the QMMSE. Thus, the generic quadratic estimator family is as follows,
\begin{align}
  g_Q(\mbf y) &= A + \mathbb{E}[\mbf x] +B\delta \mbf y + C\delta \mbf y ^{[2]}
\end{align}
where
\begin{align}
  \delta \mbf y &= \mbf y - \mathbb{E}[\mbf y] \\
  \delta \mbf y ^{[2]} &= \delta \mbf y \otimes\delta \mbf y = (\mbf y - \mathbb{E}[\mbf y]) \otimes (\mbf y - \mathbb{E}[\mbf y]) 
\end{align}
indicates the residual (or measurement deviation vector) and its square, obtained using the Kronecker product $\otimes$ for the square of vectors. 

The orthogonality principle states that, given the optimal estimator $g^*_Q(\mbf y)$, with optimal constants $A^*$, $B^*$, and $C^*$, the equation
\begin{equation}
  \mathbb{E}\bigl[(\mbf x - g^*_Q(\mbf y))\,g_Q(\mbf y)^T\bigr] = 0 \label{eq:ortho}
\end{equation}
holds for every other generic estimator. Therefore, picking the following values for the generic estimator
\begin{equation}
A =- \mathbb{E}[\mbf x] + \mbf I, \quad B = \mbf 0, \quad C = \mbf 0 
\end{equation}
where $\mbf I$ is the identity matrix, leads to
\begin{align}
  \mathbb{E}[\delta \mbf x - A^* - B^*\delta \mbf y - C^*\delta \mbf y^{[2]}] =& \;0 \nonumber, \\
  \;A^* = -C^*\,\mathrm{v}(\mbf P_{yy}) \label{eq:3}
\end{align}
where $\mathrm{v}(\mbf P_{yy})$ indicates the stack operator for matrix $\mbf P_{yy}$. Similarly, the two sets were selected
\begin{align}
A &=- \mathbb{E}[\mbf x], \quad B = \mbf I, \quad C = \mbf 0  \\
A &=- \mathbb{E}[\mbf x], \quad B = \mbf 0, \quad C = \mbf I
\end{align}
leads to 
\begin{align}
\mathbb{E}[(\delta \mbf x - A^* - B^*\delta \mbf y - C^*\delta \mbf y^{[2]})\,\delta \mbf y^T]=& \nonumber \\
  \mbf P_{xy} - B^*\mbf P_{yy} - C^*\mbf P_{y^{[2]}y} =& 0\label{eq:1}
\end{align}
and
\begin{align}
\mathbb{E}[(\delta \mbf x - A^* - B^*\delta \mbf y - C^*\delta \mbf y^{[2]})\,\delta \mbf y^{[2]T}]&= \nonumber\\ 
  \mbf P_{xy^{[2]}} - B^*\mbf P_{yy^{[2]}} + C^*(\mathrm{v}(\mbf P_{yy})\mathrm{v}(\mbf P_{yy})^T-\mbf P_{y^{[2]}y^{[2]}}) &= 0 \label{eq:2}
\end{align}
where we have defined covariances as 
\begin{align} \label{eq:set_in}
  \mbf P_{xy} &= \mathbb{E}[\delta \mbf x\,\delta \mbf y^T],\\
  \mbf P_{xy^{[2]}} &= \mathbb{E}[\delta \mbf x\,\delta \mbf y^{[2]^T}],\\
  \mbf P_{yy} &= \mathbb{E}[\delta \mbf y\,\delta \mbf y^T],\\
  \mbf P_{yy^{[2]}} &= \mathbb{E}[\delta \mbf y\,\delta\mbf  y^{[2]^T}],\\
  \mbf P_{y^{[2]}y} &= \mathbb{E}[\delta \mbf y^{[2]}\,\delta \mbf y^T],\\
  \mbf P_{y^{[2]}y^{[2]}} &= \mathbb{E}[\delta \mbf y^{[2]}\,\delta \mbf y^{[2]^T}].\label{eq:set_fin}
\end{align}
with
\begin{align}
  \delta \mbf x &= \mbf x - \mathbb{E}[\mbf x] 
\end{align}
After substituting in Eq. \eqref{eq:3}, Eqs \eqref{eq:1} and \eqref{eq:2} can be rewritten as a set of coupled equations.
\begin{equation}
  \begin{bmatrix} B^* & C^* \end{bmatrix}
  \begin{bmatrix}
    \mbf P_{yy} & \mbf P_{yy^{[2]}} \\
    \mbf P_{y^{[2]}y} & \mbf P_{y^{[2]}y^{[2]}} - \mathrm{v}(\mbf P_{yy})\mathrm{v}(\mbf P_{yy})^T
  \end{bmatrix}
  =
  \begin{bmatrix} \mbf P_{xy} \\ \mbf P_{xy^{[2]}} \end{bmatrix}^T
\end{equation}
which brings us to the solution for the optimal constants,
\begin{equation}
  \mbf K = \begin{bmatrix} B^* & C^* \end{bmatrix} =  \mbf P_{\mbf x \mathcal Y} \; \mbf P_{\mathcal Y \mathcal Y}^{-1}\label{eq:4}
\end{equation}
having defined 
\begin{align}
  \mbf P_{\mathcal X \mathcal Y} &=  \begin{bmatrix} \mbf P_{xy} & \mbf P_{xy^{[2]}} \end{bmatrix} \\
  \mbf P_{\mathcal Y \mathcal Y} &= \begin{bmatrix}
  \mbf P_{yy} & \mbf P_{yy^{[2]}}\\ \mbf P_{y^{[2]}y} & \mbf P_{y^{[2]}y^{[2]}} - \mathrm{v}(\mbf P_{yy})\mathrm{v}(\mbf P_{yy})^T
  \end{bmatrix}
\end{align}
The optimal constants $B^*$ and $C^*$ behave similarly to the classic Kalman gain, but are evaluated in its augmented form that considers the covariances and cross covariances with the square of the measurement deviation vector. 

Substituting the derived constants from Eq. \eqref{eq:3} and \eqref{eq:4} back into the definition of the quadratic estimator, it creates the optimal quadratic estimator, which is the parabolic approximation of the true MMSE,
\begin{equation}
  g^*_Q(\mbf y) = \mathbb{E}[\mbf x] + \mbf P_{\mbf x \mathcal Y}\mbf P_{\mathcal Y \mathcal Y}^{-1} 
  \begin{bmatrix}\delta \mbf y \\
          \delta \mbf y^{[2]} - \mathrm{v}(\mbf P_{yy})\end{bmatrix}
\end{equation}
The derived estimator is applied in the update step of the classic Unscented Kalman Filter (UKF) to obtain the quadratic version of their formulation, which is linear. Therefore, using the unscented transformation to evaluate the set of central moments in Eqs \eqref{eq:set_in} to \eqref{eq:set_fin}, we propose the Quadratic Unscented Kalman Filter (QUKF).

%%%%%%%%%%%%%%%%%%%%%%%%%%%%%%%%%%%%%%%%%%%%%%%%%%%%%%%%%%%%%%%%%%%%%%%%%%
\subsection{The Optimal Polynomial Estimator}
The previous process can be generalized up to any arbitrary order. We consider an optimal polynomial estimator, i.e., an estimator that is constrained to be a polynomial function of the measurement innovation but is otherwise chosen to minimize mean-square error. Concretely, instead of restricting the update to be linear in the innovation (as in the Kalman filter), we allow all monomials in the innovation up to the $N$th polynomial order.

Similarly to the quadratic estimator, we can derive an optimal estimator for the $N$th polynomial order. Following the same procedure as previously described, we  would get
\begin{equation}
  \begin{bmatrix}
    B^* \\ C^*  \\ \vdots \\ Z^*
  \end{bmatrix}^T
  \begin{bmatrix}
    \mbf P_{yy} & \mbf P_{yy^{[2]}} & ... & \mbf P_{yy^{[N]}}\\
    \mbf P_{y^{[2]}y} & \mbf P_{y^{[2]}y^{[2]}} - \mathrm{v}(\mbf P_{yy})\mathrm{v}(\mbf P_{yy})^T
      & ... & \mbf P_{y^{[2]}y^{[N]}} - \mathrm{v}(\mbf P_{yy})\mathrm{v}(\mbf P_{y^{[V]}y^{[M]}})^T\\
    \vdots & \vdots & \ddots & \vdots & \\
    \mbf P_{y^{[N]}y} & \mbf P_{y^{[N]}y^{[2]}} - \mathrm{v}(\mbf P_{y^{[V]}y^{[M]}})\mathrm{v}(\mbf P_{yy})^T &  ... & \mbf P_{y^{[N]}y^{[N]}} - \mathrm{v}(\mbf P_{y^{[V]}y^{[M]}})\mathrm{v}(\mbf P_{y^{[V]}y^{[M]}})^T
  \end{bmatrix}
  =  
  \begin{bmatrix}
    P_{xy} \\ P_{x\,y^{[2]}}  \\ \vdots \\ P_{x\,y^{[N]}}
  \end{bmatrix}^T
\end{equation}
where if $N$ is even, $M = V = N/2$ and if $N$ is odd, $M = (N+1)/2$ and $V = M-1$. The polynomial Kalman gain will be
\begin{equation}
  K \;=\; 
  \begin{bmatrix}
    B^* & C^* & D^* & ... 
  \end{bmatrix}
   \;=\; \mbf P_{\mbf x \mathcal Y} \; \mbf P_{\mathcal Y \mathcal Y}^{-1}\
\end{equation}
where 
\begin{align}
  \mbf P_{\mathcal X \mathcal Y} &= 
  \begin{bmatrix} 
    \mbf P_{xy} & \mbf P_{xy^{[2]}} & \mbf P_{xy^{[3]}} & ... & \mbf P_{xy^{[N]}}
  \end{bmatrix}
\end{align}
and
\begin{align}
  \mbf P_{\mathcal Y \mathcal Y} &= 
  \begin{bmatrix}
    \mbf P_{yy} & \mbf P_{yy^{[2]}} & ... & \mbf P_{yy^{[N]}}\\
    \mbf P_{y^{[2]}y} & \mbf P_{y^{[2]}y^{[2]}} - \mathrm{v}(\mbf P_{yy})\mathrm{v}(\mbf P_{yy})^T
      & ... & \mbf P_{y^{[2]}y^{[N]}} - \mathrm{v}(\mbf P_{yy})\mathrm{v}(\mbf P_{y^{[V]}y^{[M]}})^T\\
    \vdots & \vdots & \ddots & \vdots & \\
    \mbf P_{y^{[N]}y} & \mbf P_{y^{[N]}y^{[2]}} - \mathrm{v}(\mbf P_{y^{[V]}y^{[M]}})\mathrm{v}(\mbf P_{yy})^T &  ... & \mbf P_{y^{[N]}y^{[N]}} - \mathrm{v}(\mbf P_{y^{[V]}y^{[M]}})\mathrm{v}(\mbf P_{y^{[V]}y^{[M]}})^T
  \end{bmatrix}
\end{align}
still indicate the augmented covariance matrices for state-measurement and measurement. Thus, the optimal polynomial estimator of order N will be 
\begin{equation}
    g^*_N(\mbf y) = \mathbb{E}[\mbf x] + \mbf P_{\mbf x \mathcal Y}\mbf P_{\mathcal Y \mathcal Y}^{-1}
    \begin{bmatrix}
        \delta \mbf y \\
        \delta \mbf y^{[2]} - \mathrm{v}(\mbf P_{yy}) \\
        \delta \mbf y^{[3]} - \mathrm{v}(\mbf P_{yy^2}) \\
        \vdots \\
        \delta \mbf y^{[N]} - \mathrm{v}(\mbf P_{y^{[V]}y^{[M]}})
    \end{bmatrix}
\end{equation}

\section{High Order Central Moments}
Linear estimators require the first two central moments, i.e mean and covariance, to operate. In contrast, quadratic estimators, due to their higher order, require information up to the fourth central moment to achieve a correct and consistent estimate. Therefore, we define with 
\begin{equation}
    \mbf S_{xxx} = \mathbb{E}\bigl[(x - \hat{x}) \otimes \big((x - \hat{x})(x - \hat{x})^T\big)\bigr]
\end{equation}
the skewness of a distribution, expressed as a three-dimensional tensor, and with 
\begin{equation}
    \mbf K_{xxxx} = \mathbb{E}\bigl[\big((x - \hat{x})(x - \hat{x})\big)^T \otimes \big((x - \hat{x})(x - \hat{x})^T\big)\bigr]
\end{equation}
the kurtosis, expressed as a fourth dimensional tensor. It can be shown \cite{servadio2020nonlinear, servadio2021estimation} that for a null mean distribution, where central moments and raw moments match, $\mbf S_{xxx} = \mbf P_{x^{[2]}x} = \mbf P_{xx^{[2]}}^T $ and $\mbf K_{xxxx} = \mbf P_{x^{[2]}x^{[2]}}$. 

In the classical Kalman filter, the measurement noise is characterized solely by its second-order central moment (covariance), implicitly assuming that higher-order moments are either negligible or follow the Gaussian relationships. In contrast, the proposed polynomial estimator uses a higher-order approximation of the conditional mean and, therefore, requires a more detailed description of the measurement noise. Accurately modeling these higher-order central moments of the measurement noise leads to a richer statistical representation of the measurement distribution and enables a more accurate polynomial MMSE estimator than one based on covariance information alone.

%%%%%%%%%%%%%%%%%%%%%%%%%%%%%%%%%%%%%%%%%%%%%%%%%%%%%%%%%%%%%%%%%%%%%%%%%%
\section{The Quadratic Unscented Kalman Filter (QUKF)}\label{sec:qukf}
Consider the nonlinear equation of motion, with dynamics $f()$, affected by noise
\begin{equation}
    \mbf x_{k} = f(\mbf x_{k-1}) + \boldsymbol{\mu}_{k-1} \label{eq:sys}
\end{equation}
where $\boldsymbol{\mu}_{k-1}$ is a Gaussian process noise of zero-mean with known covariance matrix $\mbf P_{\mu\mu}$. The state PDF is  approximated as a Gaussian with known mean and covariance, which constitutes the prior for the measurement update. An observation is provided according to the measurement model $h()$,
\begin{equation}
  \mbf y = h(\mbf x) + \boldsymbol{\eta} \label{eq:meas}
\end{equation}
affected by zero-mean measurement noise with known covariance $\mbf P_{\eta\eta}$, skewness $\mbf S_{\eta\eta\eta}$, and kurtosis $\mbf K_{\eta\eta\eta\eta}$. The predicted measurement mean is indicated with $\hat{\mbf y}_k$. Given the result of the measurement $\tilde {\mbf y}$, the actual numerical value of the random value of the sensors, the residual and its square are evaluated as 
\begin{align}
    \delta \tilde{\mbf y} &= \mbf y - \hat{\mbf y}_k \\
  \delta\tilde{\mbf y}^{[2]} &= \delta \tilde{\mbf y}\otimes\delta \tilde{\mbf y} 
\end{align}
so that the quadratic update step can be performed according to 
\begin{align}
        \mbf K &= \mbf P_{x \mathcal Y}\mbf P_{\mathcal Y \mathcal Y}^{-1} \\
     \hat{\mbf x}^+_{k}&= \hat{\mbf x}^-_{k} +\mbf K
  \begin{bmatrix}\delta \mbf y \\
          \delta \mbf y^{[2]} - \mathrm{v}(\mbf P_{yy})\end{bmatrix} \label{eq:up1}\\
          \mbf P_{xx,k}^+&=\mbf P_{xx,k}^--\mbf K\mbf P_{\mathcal Y \mathcal Y}\mbf K^T \label{eq:up2}
\end{align}
where $\mbf K $ is the Kalman gain, in its augmented form, and $\hat{\mbf x}^+_{k}$ and $\mbf P_{xx,k}^+$ are the updated mean and covariance of the state distribution. Such equations are derived following the common Kalman derivation, but with the augmented system.  

The reader can notice that Eqs \eqref{eq:up1} and \eqref{eq:up2} resemble the structure of the normal Kalman update but expanded to operate with the augmented covariances that include information of the square of the measurement vector. Indeed, the state-measurement cross-covariance is evaluated block-wise as in 
\begin{equation}
    \mbf P_{x \mathcal Y} =  \begin{bmatrix} \mbf P_{xy} & \mbf P_{xy^{[2]}} \end{bmatrix}  \label{eq:pxy}
\end{equation}
In a similar pattern, the augmented measurement covariance is evaluated block-wise as in 
\begin{equation}
    \mbf P_{\mathcal Y \mathcal Y} = \begin{bmatrix}
    \mbf P_{yy} & \mbf P_{yy^{[2]}} \\
    \mbf P_{yy^{[2]}}^T & \mbf P_{y^{[2]}y^{[2]}} - \mathrm{v}(\mbf P_{yy})\mathrm{v}(\mbf P_{yy})^T
  \end{bmatrix}\label{eq:pyy}
\end{equation}
The state has been updated according to  Eqs \eqref{eq:up1} and \eqref{eq:up2} and the mean and covariance can undergo a new propagation normally. To actively complete the update step, we require a technique to obtain an accurate approximation of covariances so that the Kalman gain can be evaluated. We exploit the unscented transformation to obtain the QUKF.

The scaled unscented transformation (UT) is applied to obtain the mean and covariance of a random variable, $\mbf y$, which has a known relation, $h()$, with respect to the given initial random variable $\mbf x$, of dimension $n$. After defining the spread parameter $\alpha$, the scaling parameter $\kappa$ and the constant $\beta$, a set of  $2n+ 1$ sigma points is created from the mean of the current state and covariance: 
\begin{align}
    \lambda &= \alpha^2 (n + \kappa) - n \label{eqlambda}\\
    \mbf C_{k-1}\mbf C_{k-1}^T &=  (n+\lambda)\mbf P_{xx,k-1}^+\\
    \boldsymbol{\mathcal X}_{k-1} &= \begin{bmatrix} \hat{\mbf x}^+_{k-1}  & \hat{\mbf x}^+_{k-1} +\mbf C_{k-1}   &\hat{\mbf x}^+_{k} -\mbf C_{k-1} \end{bmatrix}
\end{align}
Each sigma point is associated with a weight given by
\begin{align}
    w_0^{(m)} &= \frac{\lambda}{n + \lambda} \\
    w_0^{(c)} &= \frac{\lambda}{n + \lambda} + (1 - \alpha^2 + \beta) \\
    w_i^{(m)} &= w_i^{(c)} = \frac{1}{2(n + \lambda)}, \quad i = 1, \dots, 2n
\end{align}
for the calculation of means, $w_i^{(m)}$, and covariances, $w_i^{(c)}$. 

The prediction step is performed by propagating, separately, each sigma point and obtaining their weighted means for the predicted estimate and covariance. 
\begin{align}
    \boldsymbol{\mathcal X}_{k}^{(i)} &= f(\boldsymbol{\mathcal X}_{k-1}^{(i)}) \quad \forall i = 0,...,2n \\
    \hat{\mbf x}^-_{k} &=  \sum^{2n}_{i=0} w^{(i)}_{m}\boldsymbol{\mathcal X}_{k}^{(i)}  \\
    \mbf P_{xx,k}^- &= \sum^{2n}_{i=0} w^{(i)}_{c} (\boldsymbol{\mathcal X}_{k}^{(i)}-\hat{\mbf x}^-_{k})(\boldsymbol{\mathcal X}_{k}^{(i)}-\hat{\mbf x}^-_{k})^T     +\mbf P_{\mu\mu}
\end{align}

The quadratic update for QUKF follows the same formulation as in Eqs \eqref{eq:up1} and \eqref{eq:up2}, with the augmented covariances reported in Eqs \eqref{eq:pxy} and \eqref{eq:pyy}, evaluated block-wise. Each entry is computed via the sigma points weighted summation. Therefore, after obtaining the transformed sigma points in the measurement space and their predicted mean,
\begin{align}
    {\mathcal Y}_{k}^{(i)} &= h({\mathcal X}_{k}^{(i)}) \quad \forall i = 0,...,2n \\
    \hat{\mbf y}_{k} &= \sum^{2n}_{i=0} w^{(i)}_{m}{\mathcal Y}_{k}^{(i)}
\end{align}
each sigma point is augmented with its Kronecker square. Thus, the deviation vectors of the state and the measurement are computed as 
\begin{align}
    \delta \boldsymbol{\mathcal X}_{k}^{(i)} &= \boldsymbol{\mathcal X}_{k}^{(i)} - \hat{\mbf x}^-_{k} \quad \forall i = 0,...,2n \\
    \delta \boldsymbol{\mathcal Y}_{k}^{(i)} &=\boldsymbol{\mathcal Y}_{k}^{(i)} -  \hat{\mbf y}_{k} \quad \forall i = 0,...,2n \\
    \delta \boldsymbol{\mathcal Y}_{k}^{[2](i)} &= \delta \boldsymbol{\mathcal Y}_{k}^{(i)} \otimes \delta \boldsymbol{\mathcal Y}_{k}^{(i)} \quad \forall i = 0,...,2n
\end{align}
This definition eases the evaluation of the covariances, which are obtained as a simple summation. The state-measurement cross covariances components are
\begin{align}
    \mbf P_{x y} &= \sum^{2n}_{i=0} w^{(i)}_{c}\delta \boldsymbol{\mathcal X}_{k}^{(i)}\delta \boldsymbol{\mathcal Y}_{k}^{(i)T} \label{eq:cv1}\\
    \mbf P_{xy^{[2]}} &= \sum^{2n}_{i=0} w^{(i)}_{c} \delta \boldsymbol{\mathcal X}_{k}^{(i)} \delta \boldsymbol{\mathcal Y}_{k}^{[2](i)T}
\end{align}
The measurement covariance is evaluated in two separate steps: first, the transformed covariances in the measurement space are computed, then the influence of the noise is added. 
\begin{align}
    \overline{\mbf P_{y y}} &= \sum^{2n}_{i=0} w^{(i)}_{c}\delta \boldsymbol{\mathcal Y}_{k}^{(i)} \delta \boldsymbol{\mathcal Y}_{k}^{(i)T}\\
    \overline{\mbf P_{yy^{[2]}}} &= \sum^{2n}_{i=0} w^{(i)}_{c} \delta\boldsymbol{\mathcal Y}_{k}^{(i)} \delta \boldsymbol{\mathcal Y}_{k}^{[2](i)T} \\
    \overline{\mbf P_{y^{[2]}y^{[2]}}} &=\sum^{2n}_{i=0} w^{(i)}_{c} \delta \boldsymbol{\mathcal Y}_{k}^{[2](i)} \delta \boldsymbol{\mathcal Y}_{k}^{[2](i)T}
\end{align}
where additive noise leads to a simple noise addition for the contribution:
\begin{align}
    \mbf P_{yy} &= \overline{\mbf P_{y y}} + \mbf P_{\eta\eta} \label{eq:cv3}\\
    \mbf P_{yy^{[2]}} &= \overline{\mbf P_{yy^{[2]}}} + \mbf S_{\eta\eta\eta}\\
    \mbf P_{y^{[2]}y^{[2]}}  &=  \overline{\mbf P_{y^{[2]}y^{[2]}}} + \mbf K_{\eta\eta\eta\eta} + \nonumber \\
    + &\overline{\mbf P_{y y}} \otimes \mbf P_{\eta\eta} + \mathrm{m}\Big( \mathrm{v}(\overline{\mbf P_{y y}}) \otimes \mbf I  \Big)\Big( \mbf P_{\eta\eta}^T \otimes \mbf I \Big) + \nonumber \\
    + & \mbf P_{\eta\eta} \otimes \overline{\mbf P_{y y}}+  
    \mathrm{m}\Big( \mathrm{v}(\mbf P_{\eta\eta}) \otimes \mbf I  \Big)\Big( \overline{\mbf P_{y y}}^T \otimes \mbf I \Big) +\nonumber\\
    + &\mathrm{v}(\overline{\mbf P_{y y}})\mathrm{v}(\mbf P_{\eta\eta})^T + \mathrm{v}(\mbf P_{\eta\eta})\mathrm{v}(\overline{\mbf P_{y y}})^T\label{eq:cv2}
\end{align}
where $\mathrm{m}()$ is the matrix operator, the inverse of the stack operator $\mathrm{v}()$, and where $\overline{\mbf P_{y y}}$ indicates the measurement covariance without any influence of noise, the actual uncertainty transformation. These expressions of the covariances are obtained by applying the expected value operator as described in Eqs \eqref{eq:set_in} to \eqref{eq:set_fin}, where full knowledge of the moments of the noise is assumed.

The QUKF algorithm is complete, where high order covariances are evaluated similarly to the classical one by weighted sigma points summations. The higher-order moments provide additional information to the filter with respect to its linear version, drastically improving the accuracy of the measurement update, which can now curve in a parabolic manner to follow the true shape of the posterior distribution. 

%%%%%%%%%%%%%%%%%%%%%%%%%%%%%%%%%%%%%%%%%%%%%%%%%%%%%%%%%%%%%%%%%%%%%%%%%%
\subsection{The Quadratic Augmented Unscented Kalman Filter (QAUKF)}
The QUKF is a filter that applies the unscented transformation to derive a solution for the QMMSE likewise the UKF is a filter that applies the unscented transformation to derive a solution for the LMMSE. In the presented application, both process and measurement noises have been assumed to be additive. Consider now the following stochastic nonlinear system, where the process noise is not additive anymore, and a measurement function where the measurement noise has a nonlinear dependence on the measurement outcome: 
\begin{align}
    \mathbf{x}_{k+1} &= f(\mathbf{x}_k,\,\boldsymbol{\mu}_k), \\
    \mathbf{y}_{k+1} &= h(\mathbf{x}_{k+1},\,\boldsymbol{\eta}_{k+1})
\end{align}
where $\boldsymbol{\mu}_k$ and $\boldsymbol\eta_{k+1}$ denote the process and the measurement noises, with
\begin{align}
\mathbb{E}[\boldsymbol{\mu}_k]=\mbf 0,\quad 
\mathbb{E}[\boldsymbol\eta_{k+1}]=\mbf 0,\quad 
\mathbb{E}[\boldsymbol{\mu}_k \boldsymbol{\mu}_k^T]=\mbf P_{\mu\mu},\quad 
\mathbb{E}[\boldsymbol\eta_{k+1} \boldsymbol\eta_{k+1}^T]=\mbf P_{\eta\eta}.
\end{align}
To embed the random disturbances directly in the unscented mapping, define the augmented state vector and covariance matrix in order to create sigma points that directly account for the influence of the noises 
\begin{align}
    \mathbf{x}_k^a
    = 
    \begin{bmatrix}
        \mathbf{x}_k & \boldsymbol{\mu}_k & \boldsymbol\eta_{k+1}
    \end{bmatrix}^T, \label{eq:71} \\
    \mathbf{\hat{x}}_k^a
    = 
    \begin{bmatrix}
        \mathbf{\hat{x}}_k & 0 & 0
    \end{bmatrix}^T, \label{eq:72}\\
    \mathbf{P}_k^a
    =
    \begin{bmatrix}
        \mathbf{P}_{xx} & 0 & 0\\
        0 & \mbf P_{\mu\mu} & 0\\
        0 & 0 & \mbf P_{\eta\eta}
    \end{bmatrix}. \label{eq:73}
\end{align}
The augmented dimension is $n_a = n_x + n_\mu + n_\eta$, such that the covariance $\mbf P_k^a$ captures both state and noise uncertainty. The unscented transformation is applied in this augmented space, creating $2n_a+1$ sigma points. The augmented sigma points are generated as
\begin{align}
    \mbf C^a_{k-1}\mbf C^{a,T}_{k-1} &=  (n_a+\lambda)\mbf P_k^a\\
    \boldsymbol{\mathcal X}_{k-1}^a &= \begin{bmatrix} \hat{\mbf x}^a_{k-1}  & \hat{\mbf x}^a_{k-1} +\mbf C_{k-1}^a   &\hat{\mbf x}^a_{k} -\mbf C_{k-1}^a \end{bmatrix}
\end{align}
after using Eq. \eqref{eqlambda} to evaluate the value of $\lambda$. Each augmented sigma point decomposes as
\begin{equation}
    \boldsymbol{\mathcal X}_{k-1}^{a(i)} =
\begin{bmatrix}
    \boldsymbol{\mathcal X}_{k-1}^{x(i)} & \boldsymbol{\mathcal X}_{k-1}^{\mu(i)} &\boldsymbol{\mathcal X}_{k-1}^{\eta(i)}
\end{bmatrix}^T
\end{equation}representing the state, process noise, and measurement noise components. Each sigma point is propagated through the process model:
\begin{align}
    \boldsymbol{\mathcal X}_{k}^{a(i)} &= f(\boldsymbol{\mathcal X}_{k-1}^{x(i)},\,\boldsymbol{\mathcal X}_{k-1}^{\mu(i)}) \quad \forall i = 0,...,2n_a 
\end{align}
yielding the predicted mean and covariance:
\begin{align}
    \hat{\mbf x}^-_{k} &=  \sum^{2n_a}_{i=0} w^{(i)}_{m}\boldsymbol{\mathcal X}_{k}^{x(i)}  \\
    \mbf P_{xx,k}^- &= \sum^{2n_a}_{i=0} w^{(i)}_{c} (\boldsymbol{\mathcal X}_{k}^{x(i)}-\hat{\mbf x}^-_{k})(\boldsymbol{\mathcal X}_{k}^{x(i)}-\hat{\mbf x}^-_{k})^T   
\end{align}
It can be noted that, with respect to the previous case, the is not a direct addition of the process noise covariance matrix as the influence of such noise has been taken care of by the additional sigma points. 

Each predicted sigma point is then passed through the measurement model
\begin{align}
    \boldsymbol{\mathcal Y}^{(i)}_k &= h(\boldsymbol{\mathcal X}_{k}^{x(i)},\,\boldsymbol{\mathcal X}_{k-1}^{\eta(i)}) \quad \forall i = 0,...,2n_a \\
    \hat{\mbf y}_{k} &= \sum^{2n_a}_{i=0} w^{(i)}_{m}\boldsymbol{\mathcal Y}_{k}^{(i)}
\end{align}
Deviations are defined similarly to the additive version of the UKF:
\begin{align}
    \delta \boldsymbol{\mathcal X}_{k}^{x(i)} &= \boldsymbol{\mathcal X}_{k}^{x(i)} - \hat{\mbf x}^{-}_{k} \quad \forall i = 0,...,2n_a \\
    \delta \boldsymbol{\mathcal Y}_{k}^{(i)} &=\boldsymbol{\mathcal Y}_{k}^{(i)} -  \hat{\mbf y}_{k} \quad \forall i = 0,...,2n_a \\
    \delta \boldsymbol{\mathcal Y}_{k}^{[2](i)} &= \delta \boldsymbol{\mathcal Y}_{k}^{(i)} \otimes \delta \boldsymbol{\mathcal Y}_{k}^{(i)} \quad \forall i = 0,...,2n_a
\end{align}
so that the state-measurement covariances can be evaluated as 
\begin{align}
    \mbf P_{x y} &= \sum^{2n_a}_{i=0} w^{(i)}_{c}\delta \boldsymbol{\mathcal X}_{k}^{x(i)}\delta \boldsymbol{\mathcal Y}_{k}^{(i)T} \\
    \mbf P_{xy^{[2]}} &= \sum^{2n_a}_{i=0} w^{(i)}_{c} \delta \boldsymbol{\mathcal X}_{k}^{x(i)} \delta \boldsymbol{\mathcal Y}_{k}^{[2](i)T}
\end{align}
while the measurement covariances are evaluated directly as
\begin{align}
    {\mbf P_{y y}} &= \sum^{2n_a}_{i=0} w^{(i)}_{c}\delta \boldsymbol{\mathcal Y}_{k}^{(i)} \delta \boldsymbol{\mathcal Y}_{k}^{(i)T}\\
    {\mbf P_{yy^{[2]}}} &= \sum^{2n_a}_{i=0} w^{(i)}_{c} \delta\boldsymbol{\mathcal Y}_{k}^{(i)} \delta \boldsymbol{\mathcal Y}_{k}^{[2](i)T} \\
    {\mbf P_{y^{[2]}y^{[2]}}} &=\sum^{2n_a}_{i=0} w^{(i)}_{c} \delta \boldsymbol{\mathcal Y}_{k}^{[2](i)} \delta \boldsymbol{\mathcal Y}_{k}^{[2](i)T}
\end{align}
which implicitly contains the measurement-noise covariance $\mbf P_{\eta\eta}$ since $\boldsymbol \eta_k$ is part of $\mbf x_k^a$. 

By using this augmented version of the unscented transformation, where the influence of the noise is already accounted for by the larger set of sigma points, we do not need to adjust the covariance for the measurement noise contribution, as shown in Eqs. \eqref{eq:cv3} to \eqref{eq:cv2}. Thus, the derivation of each covariance block for the quadratic update becomes extremely trivial and easy to compute, as the full algorithm reduces to simple summations. However, it is important to underline that this implementation works well for Gaussian noises, while the previous one is more robust to any noise color, since high-order moments are directly included by their value rather than approximated by the unscented transformation, which implements a Gaussian assumption during the creation of the augmented set of sigma points. Nevertheless, this implementation is convenient for most applications, as all noise effects are inherently included within the expectations above, yielding a compact and self-consistent formulation suitable for the Quadratic Augmented UKF (QAUKF). Indeed, the correction step of the QAUKF follows the QMMSE update, likewise the QUKF, with the only difference being how the covariances are evaluated. 

%%%%%%%%%%%%%%%%%%%%%%%%%%%%%%%%%%%%%%%%%%%%%%%%%%%%%%%%%%%%%%%%%%%%%%%%%%
\section{Generalization of Quadratic Updates}
A quick consideration regarding the application and generalization of the quadratic update and the QMMSE approximation is due. The proposed quadratic update fits any technique that evaluate central moments and expectations, and it can be applied to other linear estimators, such as the Cubature Kalman Filter or the Ensemble Kalman Filter, where it changes how Eqs \eqref{eq:cv1} to \eqref{eq:cv2} are evaluated, while keeping the same structure for the augmented squaring of the measurement update. 

For example, in the preliminary version of this work, a Quadratic Extended Kalman Filter (QEKF) has been introduced \cite{servadio2025quadratic_article}. The two filters (QEKF and QUKF) have the same structure, similar to how both the EKF and UKF follow the Kalman formulation. They create an augmented system that follows the same procedure regardless of the uncertainty transformation technique selected: linearization for the QEKF and unscented transformation for the QUKF.

That is, the quadratic update structure can be utilized with any preferred moments propagation technique, e.g., quadrature or state transition tensors (STT), as long as the evaluation of the high-order central moments is accurate.

%%%%%%%%%%%%%%%%%%%%%%%%%%%%%%%%%%%%%%%%%%%%%%%%%%%%%%%%%%%%%%%%%%%%%%%%%%
\section{Generalization of the UKF to Arbitrary Order}
The QUKF and QAUKF derived so far are a quadratic approximation of the true QMMSE, which indicates the optimal quadratic estimator in a minimum mean square error sense. As the LMMSE is the linear approximation of the true MMSE, and the QMMSE is the parabolic approximation, increasing the order of the polynomial estimator will yield more accurate results, as the polynomial function better resembles the shape of the true MMSE, which can be any generic function, as shown in \cite{jaifsurvey}. Therefore, the polynomial estimator can be increased to better match the true MMSE; for example, a third-order polynomial yields the Cubic MMSE (CMMSE). 

Thus, the presented QAUKF can be expanded to the Polynomial Augmented Unscented Kalman Filter (PAUKF), where the estimator order is arbitrary high. Using the same assumptions and initial conditions as in Section \ref{sec:qukf}, consider 
\begin{align}
  \delta{\mbf y}^{[N]} &= \delta {\mbf y}\otimes\delta {\mbf y}\otimes\delta {\mbf y}\otimes..... (N \; \text{times})
\end{align}
the $N$th order residual, so that the updated mean becomes
\begin{align}
    \hat{\mbf x}^+_{k}&= \hat{\mbf x}^-_{k} +\mbf K
    \begin{bmatrix}\delta \mbf y \\
          \delta \mbf y^{[2]} - \mathrm{v}(\mbf P_{yy}) \\
          \delta \mbf y^{[3]} - \mathrm{v}(\mbf P_{yy^[2]}) \\
          \vdots\\
          \delta \mbf y^{[N]} - \mathrm{v}(\mbf P_{y^{[V]}y^{[M]}})
    \end{bmatrix}
\end{align}
where the augmented Kalman gain retains the form 
\begin{equation}
    \mbf K = \mbf P_{\mbf x \mathcal Y} \; \mbf P_{\mathcal Y \mathcal Y}^{-1}
\end{equation}
but in their expanded form, where the state-measurement cross-covariance is evaluated as
\begin{equation}
    \mbf P_{\mbf x \mathcal Y} =  \begin{bmatrix} \mbf P_{xy} & \mbf P_{xy^{[2]}} & \mbf P_{xy^{[3]}} & ..... & \mbf P_{xy^{[N]}}\end{bmatrix}
\end{equation}
while the augmented measurement covariance is
\begin{align}
  \mbf P_{\mathcal Y \mathcal Y} &= 
  \begin{bmatrix}
    \mbf P_{yy} & \mbf P_{yy^{[2]}} &  ... & \mbf P_{yy^{[N]}}\\
    \mbf P_{y^{[2]}y} & \mbf P_{y^{[2]}y^{[2]}} - \mathrm{v}(\mbf P_{yy})\mathrm{v}(\mbf P_{yy})^T
      &  ... & \mbf P_{y^{[2]}y^{[N]}} - \mathrm{v}(\mbf P_{yy})\mathrm{v}(\mbf P_{y^{[V]}y^{[M]}})^T\\
    \vdots & \vdots  & \ddots & \vdots & \\
    \mbf P_{y^{[N]}y} & \mbf P_{y^{[N]}y^{[2]}} - \mathrm{v}(\mbf P_{y^{[V]}y^{[M]}})\mathrm{v}(\mbf P_{yy})^T &  ... & \mbf P_{y^{[N]}y^{[N]}} - \mathrm{v}(\mbf P_{y^{[V]}y^{[M]}})\mathrm{v}(\mbf P_{y^{[V]}y^{[M]}})^T.
  \end{bmatrix}
\end{align}
Each block-wise entry of the covariances is evaluated directly through the sigma points summation using the unscented transformation, in the following way:
\begin{align}
    \mbf P_{xy^{[N]}} &= \sum^{2n_a}_{i=0} w^{(i)}_{c}\delta \boldsymbol{\mathcal X}_{k}^{(i)}\delta \boldsymbol{\mathcal Y}_{k}^{[N](i)T} \\
    \mbf P_{y^{[V]}y^{[M]}} &= \sum^{2n_a}_{i=0} w^{(i)}_{c} \delta \boldsymbol{\mathcal Y}_{k}^{[V](i)} \delta \boldsymbol{\mathcal Y}_{k}^{[M](i)T}
\end{align}
Eventually, a higher-order polynomial for the representation of the MMSE will stop being beneficial, as the new added terms will bring fewer beneficial contributions, the same way (for example)  that the difference between the 11th and 12th order polynomial approximation of a nonlinear function shows almost no improvement when compared to the difference in approximation between the 2nd and 3rd order. In a similar way, evaluating extremely high-order central moments can be inaccurate with the limited number of sigma points generated by the classic unscented transformation. For this reason, we expanded the PAUKF using the Conjugate Unscented Transformation (CUT) to evaluate expectations more accurately and with a higher fidelity.

%%%%%%%%%%%%%%%%%%%%%%%%%%%%%%%%%%%%%%%%%%%%%%%%%%%%%%%%%%%%%%%%%%%%%%%%%%%%
%%%%%%%%%%%%%%%%%%%%%%%%%%%%%%%%%%%%%%%%%%%%%%%%%%%%%%%%%%%%%%%%%%%%%%%%%%%%
%%%%%%%%%%%%%%%%%%%%%%%%%%%%%%%%%%%%%%%%%%%%%%%%%%%%%%%%%%%%%%%%%%%%%%%%%%%%
\section{The Conjugate Unscented Transformation for the Polynomial Update}\label{sec:cut}
The Conjugate Unscented Transformation (CUT) is a high-order sigma-point method for approximating multivariate Gaussian expectations, generalizing the classical Unscented Transformation by systematically increasing the polynomial degree of exactness. Instead of using a small fixed set of symmetric sigma points, CUT constructs conjugate point sets by tensoring one-dimensional Gaussian quadrature rules along conjugate directions, yielding formulated chosen samples that exactly integrate multivariate polynomials up to a prescribed order, as shown in the work of Adurthi et al. in \cite{adurthi2018cut_applications}. For a given order, the associated CUT rule matches all mixed moments of the underlying distribution, enabling accurate propagation and transformation of not only mean and covariance, but also higher-order central moments required by polynomial estimators. That is, when increasing the polynomial update of the PAUKF to high orders, it is expected for the UT to fail in the expected values computations. The introduction of the CUT will straighten the filtering technique, effectively allowing an arbitrary high polynomial estimator order.

%%%%%%%%%%%%%%%%%%%%%%%%%%%%%%%%%%%%%%%%%%%%%%%%%%%%%%%%%%%%%%%%%%%%%%%%%%%%
\subsection{The Conjugate Unscented Transformation}
Nonlinear moment propagation requires evaluating expectations of the form
\begin{equation}
\mathbb{E}\!\big[f(\mbf{x})\big] \;=\; \int_{\mathbb{R}^n} f(\mbf{x})\,p(\mbf{x})\,\mathrm{d}\mbf{x},
\end{equation}
which rarely admit closed forms beyond special cases. A long line of numerical integration methods approximates such integrals with deterministic point-weight rules, from classical Gaussian quadrature and fully symmetric cubature to sparse grids \cite{stroud1971multiple, stroud1966gaussian, gerstner1998sparse, stroud1960degree2, mcnamee1967fully}. Within Bayesian estimation, this perspective underlies sigma-point filters: the UT and its variants replace the integral with a small set of weighted evaluations that reproduce low-order moments exactly and approximate higher-order ones \cite{julier2000new, wu2006numerical, julier2002reduced}. The UT family achieves third-degree exactness with $\mathcal{O}(n)$ points, offering a favorable cost-accuracy trade for many problems; however, third-degree rules can miscalibrate covariances in strongly nonlinear regimes or in the presence of non-Gaussian noise, motivating higher-degree but still scalable alternatives.

The Conjugate Unscented Transformation (CUT) constructs symmetric sets of higher-degree sigma-points from conjugate tuples (signed permutations of unit vectors aligned with the axis) so that all odd central moments vanish by symmetry and targeted even moments are matched by design, as shown in \cite{adurthi2012cut, adurthi2012ms, adurthi2018cut_applications}.

The CUT rules (e.g., CUT4, CUT6, CUT8) enforce fourth-, sixth-, or eighth-order moment constraints for a given prior family while keeping the number of points polynomial in dimension (typically $\mathcal{O}(n)$, $\mathcal{O}(n^2)$, or $\mathcal{O}(n^3)$ depending on the degree), thus avoiding the exponential growth of Gauss-Hermite formulas while exceeding the accuracy of third-degree UT \cite{stroud1971multiple, stroud1966gaussian, julier2000new, wu2006numerical, adurthi2012cut, adurthi2012ms, adurthi2018cut_applications}. %The CUT has been demonstrated in estimation and control \cite{adurthi2018cut_applications}, in high-fidelity space applications such as conjunction analysis \cite{adurthi2015cut}, and extended to non-Gaussian priors, including uniform densities \cite{adurthi2013cut_uniform}. 
%Related methodologies such as polynomial chaos expansions and Gaussian mixture representations offer complementary uncertainty-propagation tools with different accuracy-cost profiles \cite{madankan2014hazard, xiu2002wiener, madankan2013pcbayes, terejanu2008gmm, terejanu2011agsf}.

The CUT generates a set of deterministic sigma points to match the central moments of a Gaussian distribution. Considering the state random variable in terms of its scaled deviation with respect to the mean, variable $\boldsymbol \xi$, such that 
\begin{equation}
    \mbf{x} = \mbf{\hat x} + \mbf{C}\,\boldsymbol{\xi}, \label{eq:map} 
\end{equation}
with $\mbf{P} = \mbf{C}\mbf{C}^{T}$ holds $\boldsymbol{\xi}\sim\mathcal N(\mbf 0,\mbf I)$, the integral evaluation requested by the expected value operator is approximated in CUT by a weighted sum
\begin{equation}
\mathbb{E}\!\big[f(\mbf{x})\big] \;=\; \int_{\mathbb{R}^n} f(\mbf{x})\,p(\mbf{x})\,\mathrm{d}\mbf{x} \approx \sum_i w^{(i)} f(\mbf{\hat x}  + \mbf C\boldsymbol{\xi}^{(i)})
\end{equation}
where the sigma points $\boldsymbol{\xi}^{(i)}$ and weights $w^{(i)}$ are selected such that this quadrature rule is exact for multivariate polynomials up to a prescribed degree under the Gaussian measure. That is, CUT generates weights and sigma points to match Gaussian moments. Exactness of the cubature rule for polynomial integrands translates into a hierarchy of moment-matching constraints on the sigma points. The zeroth-order constraint 
\begin{equation}
\sum_i w^{(i)} = 1
\end{equation}
ensures exact integration of constants. First-order exactness requires the weighted mean of the sigma points to vanish,
\begin{equation}
\sum_i w^{(i)} \boldsymbol{\xi}^{(i)} = \mbf{0}
\end{equation}
while the second-order exactness enforces 
\begin{equation}
\sum_i w^{(i)} \boldsymbol{\xi}^{(i)} \boldsymbol{\xi}^{(i)\tr} = \mbf{I},
\end{equation}
which guarantees exact reproduction of the covariance. A defining feature of CUT is the use of conjugate-symmetric sigma-point sets, which automatically eliminate all odd-order central moments, yielding
\begin{equation}
\sum_i w^{(i)} \boldsymbol{\xi}^{(i)}\!\otimes\!\boldsymbol{\xi}^{(i)}\!\otimes\!\boldsymbol{\xi}^{(i)} = \mbf{0}
\end{equation}
This symmetry is crucial for suppressing third-order truncation errors in nonlinear propagation. To achieve fourth-order accuracy, CUT enforces exact matching of the fourth central moments of a standard multivariate Gaussian. This is expressed compactly by the tensor identity
\begin{align}
    \sum_i w^{(i)} \,\boldsymbol\xi^{(i)}_p \boldsymbol\xi^{(i)}_q \boldsymbol\xi^{(i)}_r \boldsymbol\xi^{(i)}_s 
    = \delta_{pq}\delta_{rs}+\delta_{pr}\delta_{qs}+\delta_{ps}\delta_{qr}, && \forall\, p,q,r,s\in\{1,\dots,n\}.
\end{align}
which is the same as the kurtosis evaluation of a Gaussian using Isserlis's formulation \cite{isserlis1918formula}, as shown in previous works \cite{servadio2025quadratic_article}. For example, in scalar form, these constraints become
\begin{align}
\sum_i w^{(i)} \big(\xi_k^{(i)}\big)^{4} = 3, \quad
\sum_i w^{(i)} \big(\xi_k^{(i)}\big)^{2}\big(\xi_j^{(i)}\big)^{2} = 1 \quad (k\neq j),
\end{align}
corresponding to kurtosis and cross-kurtosis of the multivariate Gaussian \cite{adurthi2012cut, adurthi2012ms, adurthi2018cut_applications}, concluding the constraint equations set to implement a 4th order CUT. Importantly, these moment constraint equations depend only on the degree of exactness, not on the state dimension $n$. Consequently, once a CUT rule of a given order is derived, the associated radii and weights can be reused for arbitrary dimensions. The sigma points in the original state space are then obtained via the affine mapping Eq. \eqref{eq:map}.

\begin{figure}  [htbp!]
    \centering
    \begin{subfigure}{0.4\textwidth}
        \centering
        \includegraphics[width=\textwidth]{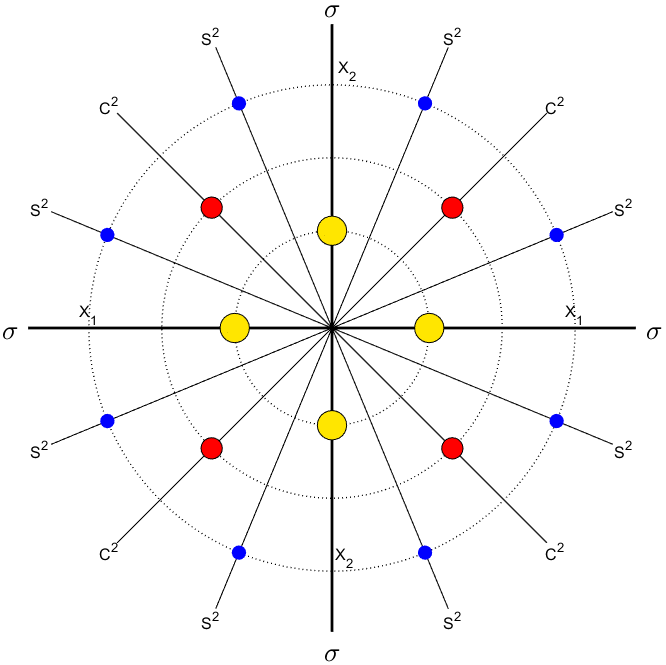}
        \caption{}
        \label{fig:CUT2D}
    \end{subfigure}
    \hfill
    \begin{subfigure}{0.4\textwidth}
        \centering
        \includegraphics[width=\textwidth]{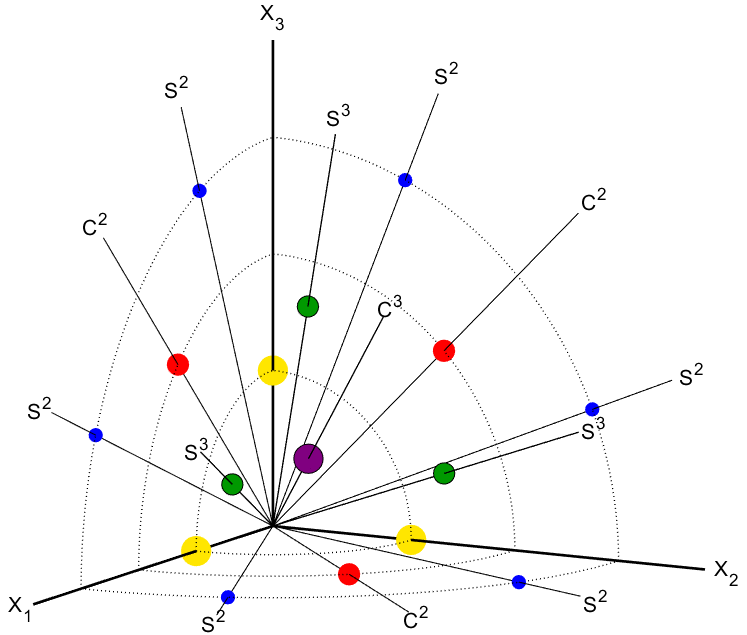}
        \caption{}
        \label{fig:CUT3D}
    \end{subfigure}
    \caption{Symmetric set of CUT points and axes in (a) 2D space and (b) 3D first octant space}
    \label{fig:CUT_pts}
\end{figure}
Figure \ref{fig:CUT_pts} gives a visual representation of the set of conjugate sigma points in the bi-dimensional (a) and three-dimensional (b) case. Increasing the CUT $c$th-order physically means cutting the state space by an additional segment, thereby creating a larger set of conjugate sigma points. Therefore, CUT would reduce to the standard UT if an order $c=2$ were selected, as the moment-matching constraint equations would stop at the covariance level, yielding only the standard sigma points for the mean and covariance (yellow dots in the figure). The larger sigma-point set, designed to match higher moments, provides a more accurate evaluation of the expectations required by the polynomial update.  

Given the nonlinear dynamics system of Eq. \eqref{eq:sys} and the nonlinear measurement model of Eq. \eqref{eq:meas}, the CUT can substitute the standard UT in the evaluation of the predicted means and covariances. Indeed, after creating the set of conjugate sigma points, whose number depends on the conjugate transformation order, the measurement mean and covariance are evaluated as deterministic summation, likewise in the standard UT. 

Because CUT reproduces higher-order moments of the prior, these weighted reductions incur smaller truncation errors than third-degree rules (standard UT), improving covariance calibration and residual consistency in strongly nonlinear mappings \cite{julier2000new, wu2006numerical, adurthi2012cut, adurthi2012ms, adurthi2018cut_applications}. Once again, when process and measurement noises are not additive, the standard augmented-state is used (Eq. \eqref{eq:71}), and the CUT generates the corresponding number of conjugate sigma points to account for the noise: $n_{aCUT}$.

For CUT4 with single- and pair-axis conjugate tuples, the sigma-point count scales as $\mathcal{O}\!\big(n+n(n-1)\big)=\mathcal{O}(n^2)$; including triple-axis tuples for CUT6/CUT8 yields $\mathcal{O}(n^3)$ growth. These counts remain polynomial in $n$, markedly below Gauss-Hermite and comparable to sparse-grid rules at similar accuracy levels \cite{stroud1966gaussian, gerstner1998sparse}. Conjugate symmetry guaranties the exact cancellation of all odd central moments and typically admits nonnegative weights, which helps to maintain positive-definite covariance updates and improves numerical stability in square-root implementations \cite{adurthi2012cut, adurthi2012ms, adurthi2018cut_applications}. When the prior is non-Gaussian, specialized CUT rules can be designed. For example, for uniform densities, by enforcing the appropriate moment conditions for that prior \cite{adurthi2013cut_uniform}.

CUT requires (i) a one-time solution of the scalar moment equations to obtain radii and weights for the chosen degree and prior family; (ii) a square-root factorization at each step; and (iii) vectorized propagation of all sigma points through $f()$ and $h()$. The resulting means, covariances, and cross-covariances are computed by the same weighted sums as UT, making replacement straightforward in PAUKF. 

%%%%%%%%%%%%%%%%%%%%%%%%%%%%%%%%%%%%%%%%%%%%%%%%%%%%%%%%%%%%%%%%%%%%%%%%%%
\subsection{The Polynomial Augmented Conjugate Unscented Kalman Filter (PACUKF-$c$)}
Building upon the augmented unscented transformation, the Polynomial Augmented Conjugate Unscented Kalman Filter (PACUKF-$c$) extends the PAUKF by replacing the classical sigma set with the CUT$c$ point set of order $c\!\in\!\{4,6,8\}$. This construction preserves moment matching up to order~$c$ while retaining the augmented formulation that embeds both process and measurement noises.

As before, define the augmented state using Eqs. \eqref{eq:71}, \eqref{eq:72}, and \eqref{eq:73}. Let $\{(\boldsymbol{\xi}^{(i)},\,\boldsymbol\mu^{(i)})\}_{i=0}^{2n_{aCUT}}$ be the set of conjugate points of order~$c$ in $\mathbb{R}^{n_a}$ that satisfies the moment constraints, so that statistical moments up to order~$c$ are matched exactly, e.g., $c=6 \Rightarrow$ CUT6 (sixth-order).

Therefore, the PAUKF is improved to the PACUKF-$c$ by replacing UT with CUT$c$, where the only difference from the previous derivation is the number of deterministic samples, namely, a total of $2n_{aCUT}+1$ conjugate sigma points. After creating the set of conjugate augmented sigma points $\boldsymbol{\mathcal{X}}^{a(i)}$ with the desired order of CUT$c$, the PACUKF-$c$ follows the same filtering structure, here summarized:
\begin{align}
    \boldsymbol{\mathcal X}_{k}^{a(i)} &= f(\boldsymbol{\mathcal X}_{k-1}^{x(i)},\,\boldsymbol{\mathcal X}_{k-1}^{\mu(i)}) \quad \forall i = 0,...,2n_{aCUT} \\
    \hat{\mbf x}^-_{k} &=  \sum^{2n_{aCUT}}_{i=0} w^{(i)}_{m}\boldsymbol{\mathcal X}_{k}^{x(i)}  \\
    \mbf P_{xx,k}^- &= \sum^{2n_{aCUT}}_{i=0} w^{(i)}_{c} (\boldsymbol{\mathcal X}_{k}^{x(i)}-\hat{\mbf x}^-_{k})(\boldsymbol{\mathcal X}_{k}^{x(i)}-\hat{\mbf x}^-_{k})^T\\
    \boldsymbol{\mathcal Y}^{(i)}_k &= h(\boldsymbol{\mathcal X}_{k}^{x(i)},\,\boldsymbol{\mathcal X}_{k-1}^{\eta(i)}) \quad \forall i = 0,...,2n_{aCUT} \\
    \hat{\mbf y}_{k} &= \sum^{2n_{aCUT}}_{i=0} w^{(i)}_{m}\boldsymbol{\mathcal Y}_{k}^{(i)}\\
    \mbf P_{xy^{[J]}} &= \sum^{2n_{aCUT}}_{i=0} w^{(i)}_{c}\delta \boldsymbol{\mathcal X}_{k}^{x(i)}\delta \boldsymbol{\mathcal Y}_{k}^{[J](i)T} \quad \forall J = 1,...,N\\
    \mbf P_{y^{[J]}y^{[S]}} &= \sum^{2n_{aCUT}}_{i=0} w^{(i)}_{c} \delta \boldsymbol{\mathcal Y}_{k}^{[J](i)} \delta \boldsymbol{\mathcal Y}_{k}^{[S](i)T}\quad \forall J,S = 1,...,N \\
    \mbf K &= \mbf P_{x \mathcal Y}\mbf P_{\mathcal Y \mathcal Y}^{-1} \\
     \hat{\mbf x}^+_{k}&= \hat{\mbf x}^-_{k} +\mbf K \begin{bmatrix}\delta \mbf y, & \dots, & 
    \delta \mbf y^{[N]} - \mathrm{v}(\mbf P_{y^{[V]}y^{[M]}})\end{bmatrix}^T \\
    \mbf P_{xx,k}^+&=\mbf P_{xx,k}^--\mbf K\mbf P_{\mathcal Y \mathcal Y}\mbf K^T 
\end{align}
This PACUKF-$c$ algorithm is a polynomial update filter of order $N$ that embeds a conjugate unscented transformation of order $c$ to approximate central moments.

\section{Numerical Applications}
The proposed algorithms have been applied to three different applications: firstly, the newly developed update techniques have been used in a simple scalar toy problem to visually highlight the benefits of the quadratic-CUT approach; secondly, they were tested with relative navigation in space application using Clohessy-Wiltshire equations with non-Gaussian measurement noises. Finally, the methodology was pitted against the Circular Restricted 3 Body Problem with Gaussian measurement noises.

As multiple filters with a different set of options have been covered, their nomenclature is here summarized. The general shorthand filter name is in the form of the form ``[estimator order][augmentation][method][Kalman Filter]-$c$", where:
\begin{itemize}
    \item {[estimation order]} $\in$ \{[$\cdot$],[Q],[C],[P]\} indicates the order of the polynomial update, with [$\cdot$] for linear (e.g. UKF), [Q] for quadratic (e.g. QUKF), [C] for cubic (e.g. CUKF), and [P] for the general unspecified order (e.g., PUKF). 
    \item {[augmentation]} $\in$ \{[$\cdot$],[A]\} denotes if the UT and CUT have been applied in their augment form, i.e., [A], or in their additive noise form, i.e., [$\cdot$].
    \item {[method]} $\in$ \{[U],[CU]\} specifies the selected methodology for the evaluation of expectations: [U] for UT, and [CU] for CUT.
    \item $c$ indicates the order of the conjugate unscented transformation in the case that CUT has been selected.
\end{itemize}
For example, the selection CACUKF-6 indicates a cubic estimator embedded with the augmented conjugate transformation of sixth order. As is evident, each name combination builds upon the standard UKF.

A quick consideration is due. As the standard UT is evaluated to match the covariance of a Gaussian distribution, the lowest order considered in the CUT is $c=4$, as the second order CUT reduces to the UT, i.e., CUT2=UT, and odd moments have null contribution in the Gaussian case, e.g., CUT7=CUT6. That is, the filter QAUKF is identical to QACUKF-2.

%%%%%%%%%%%%%%%%%%%%%%%%%%%%%%%%%%%%%%%%%%%%%%%%%%%%%%%%%%%%%%%%%%%%%%%%%
\subsection{Scalar Problem}
A scalar problem is formulated to verify the improvements of the quadratic update, enabling nonlinear estimation, over any linear filtering techniques, such as the UKF. The problem will show how the cubic representation (CMMSE) better approximates the true MMSE function, especially when compared to its linear counterpart, the LMMSE, and its quadratic counterpart, the QMMSE.

Define a prior state $x \sim \mathcal{N}(1,0.05)$ and a measurement
\begin{align}
y = \mathrm{arctan}(x) + \eta
\end{align}
where $\eta \sim \mathcal{N}(0,0.01)$ is the measurement noise and $\mathcal{N}$ indicates a Gaussian distribution.

\begin{figure}  [htbp!]
    \centering
    \includegraphics[width=1\linewidth]{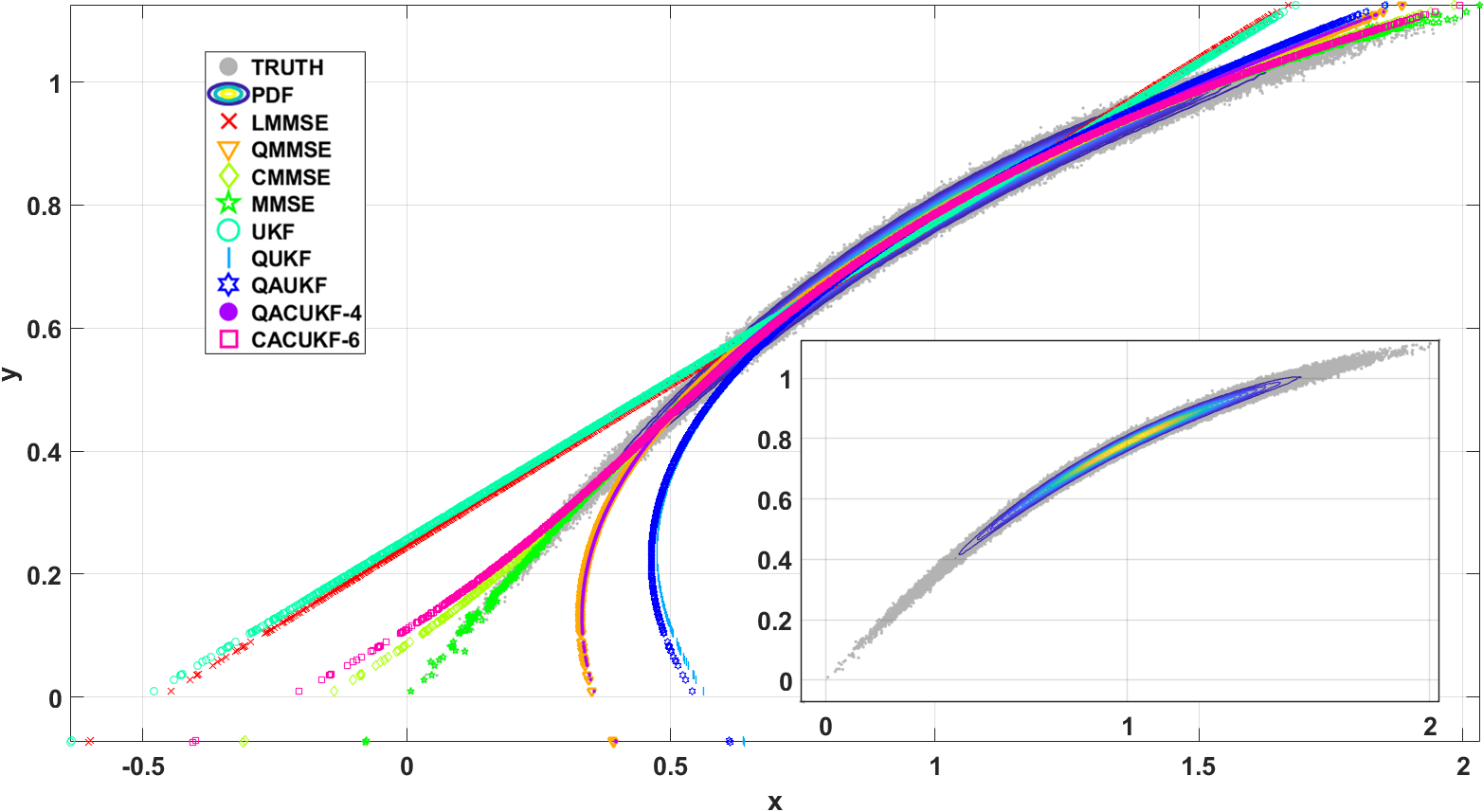}
    \caption{Representation of the state-measurement joint distribution and of the relative estimates from different estimators.}
    \label{fig:post_est}
\end{figure}

We then simulate the true joint distribution of $x$ and $y$ using $10^6$ samples, reported in Fig. \ref{fig:post_est}. The optimal (nonlinear) MMSE is the conditional mean, which visually is the curved line that divides in half the distribution of $y$ (horizontal spread of gray points) for each value of $x$. Starting from a Gaussian distribution, the spread of the points follows the joint PDF (thus, the posterior) shape due to the nonlinearity of the measurement equation. The figure also shows different estimators in various colors: UKF, QUKF, QAUKF, QACUKF-4, and CACUKF-6. The UKF (cyan) is the  straight line with the empty circle marker whose slope is the Kalman gain. A more accurate transformation of central moments leads to a better prediction in the measurement space and, therefore, to a more reliable Kalman gain. Conceptually, the optimal Kalman gain obtainable is represented by the red line, which represents the true LMMSE evaluated directly from the particles. Linear estimators, regardless of the technique that can be used to approximate moments, aim asymptotically at this line.
In contrast, quadratic estimators have the capability of obtaining a curved function. In the figure, the QUKF calculates $P_{xy^{[2]}}$ directly from the sigma points, giving a non-null skewed behavior. The result is an estimator line that follows a parabolic trend in the variable $y$, as shown by the blue line. The QUKF estimate turns and follows the curved shape of the posterior distribution, achieving a more accurate estimate. The QUKF curved behavior improves both the representation of the tails of the distribution and the region around the mode of the posterior PDF. The QAUKF, denoted by the blue star, follows the same logic but shows a little better improvement compared to QUKF due to its noise augmented nature in its propagation.

The figure also shows the QACUKF-4, denoted by the purple full circle, which follows the curve of the PDF more accurately compared to the QUKF. This is due to the introduction of quadrature points generated by CUT. For every dimension, a combination of symmetric axes is selected from which points on the same set are obtained. Due to this, the accuracy of the QACUKF-4 is greatly improved. The curve of QACUKF-4 is observed to be nearly identical to the slope of QMMSE which is represented as an orange inverted triangle. Once again, the QMSSE is the optimal quadratic estimator, obtained directly form the points of the PDF. Lastly, we have the CACUKF-6 line denoted by the pink square. This line resembles the distribution the most when compared to other filters. This is mainly due to its higher order moments in the update as well as the higher CUT order, giving us more quadrature points to propagate from.

\begin{figure}  [htbp!]
    \centering
    \includegraphics[width=.8\linewidth]{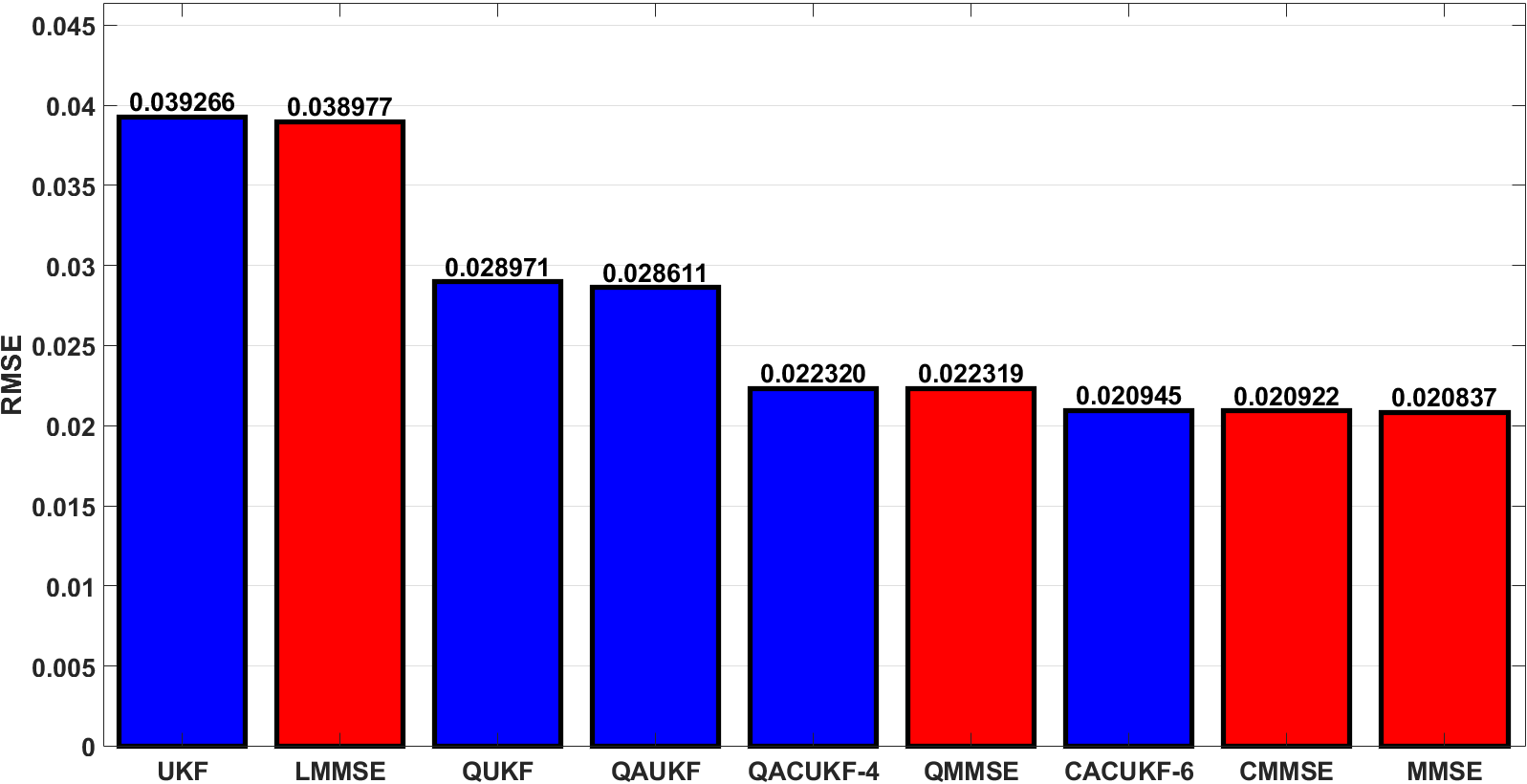}
    \caption{RMSE among selected estimators}
    \label{fig:rmse}
\end{figure}

The advantages of the quadratic approximation of the MMSE over the linear counterpart can be appreciated in Fig. \ref{fig:rmse}, where the root mean square error has been evaluated as
\begin{equation}
    RMSE = \sqrt{ \sum_{i=1}^{N_{MC}} (\hat{x}_i - x_T)^2}
\end{equation}
where $x_T$ is the true value of $x$ and $N_{MC}$ is the number of points representing the distribution. The filters have blue bars, while the MMSE bars are red to better distinguish between them. The LMMSE bar represents the best possible linear fit of the particles that make up the posterior distribution. The UKF is an accurate linear estimator whose performance is close to best possible obtainable LMMSE. Regardless of the techniques used, no linear estimator can break the accuracy threshold represented by the LMMSE bar. The QUKF is more accurate than the UKF, improving accuracy drastically thanks to the quadratic update being more reliable than mere linearization of the MMSE. Next, by augmenting the propagation with noise, the QAUKF does a slightly better improvement compared to the QUKF. However, it pales in comparison to the performance of the QACUKF-4 filter. This filter takes advantage of the extra sigma points provided by the CUT4 and gives an RMSE value which is really close to the QMMSE RMSE, thereby showing it's effectiveness even in only second order filters. Indeed, different quadratic estimators differs on how well they approximate the high order central moments, providing more accurate Kalman gain constants. Regardless on the level of accuracy, no quadratic estimator can achieve a lower RMSE than the bar represented by the true QMMSE, as it is fitted directly from the simulated points. 

To prove that this is not a coincidence, the third order filters are implemented. The CACUKF-6 shows us that even with cubic terms, the filter is neck in neck with its corresponding MMSE. However, the CACUKF-6 cannot exceed the limit given by the optimal CMMSE, which represents one of the most accurate polynomial estimators achievable.
On the other hand, the CACUKF-6 is able to better follow the shape of the distribution, and its error is much lower than the UKF, with an improvement of around 25\% in accuracy. Ref \cite{servadio2020nonlinear} showed that, conceptually, by increasing the order of the polynomial update higher, the accuracy keeps improving, until asymptotically it reaches the true MMSE. The PACUKF-$c$ derived in this paper enhances the use of the unscented transformation (and CUT) to obtain a polynomial representation of the true MMSE that outperforms linear estimators.

%%%%%%%%%%%%%%%%%%%%%%%%%%%%%%%%%%%%%%%%%%%%%%%%%%%%%%%%%%%%%%%%%%%%%%
\subsection{Clohessy-Wiltshire Relative Motion}
The Clohessy-Wiltshire (CW) equations, also known as Hill's equations, describe the relative motion of a chaser spacecraft with respect to a target spacecraft in a circular orbit. Assuming the target is in a circular orbit and a local-vertical-local-horizontal (LVLH) coordinate frame centered on the target, the linearized equations of motion for the relative position vector $\mathbf{r} = [x, y, z]^T$ are given by:
\begin{align}
    \ddot{x} &= 2\alpha\dot{y} + 3\alpha^2x \\
    \ddot{y} &= -2\alpha\dot{x} \\
    \ddot{z} &= -\alpha^2z
\end{align}
where $\alpha$ is the mean motion of the chief's circular orbit, defined as \( \alpha = \sqrt{\mu / a^3} \), with $\mu$ the Earth's standard gravitational parameter, and $a$ the semi-major axis of the chief's orbit, assumed to be 7000 km. These equations assume linearized dynamics, valid for small separations relative to the orbital radius. The initial relative PDF is assumed to be a Gaussian distribution with mean $\mathbf{x}_0 = [
\mathbf{r}_0 \
\mathbf{v}_0 ]^T$ 
given by
\begin{align}
\mathbf{r}_0 &= 
\begin{bmatrix}
2 &
10 &
-3.5 
\end{bmatrix}^T \quad \text{km}\\
\mathbf{v}_0 &= 
\begin{bmatrix}
0.01 &
-0.005 &
0.0005
\end{bmatrix} ^T \quad \text{km/s}
\end{align}
and covariance \( \mathbf{P}_0 \) defined as:
\begin{equation}
\mathbf{P}_0 = \mathrm{blkdiag}\left(10^{-4}\mathbf{I}_3\text{km}^2,10^{-9}\mathbf{I}_3\text{km/s}^2\right)
\end{equation} 
where $\mathrm{blkdiag}()$ indicates the diagonal block operator and $\mathbf{I}_3$ a 3x3 identity matrix. 

Relative angle measurements are acquired by the chief every minute for 3 hours according to 
\begin{align}
y_1 &= \mathrm{arctan}(y/x) + \eta_1 \\
y_2 &= \mathrm{arcsin}\left(\dfrac{z}{\sqrt{x^2+y^2+z^2}}\right) + \eta_2
\end{align}
where $\eta_i$ is a non-Gaussian noise of zero mean, whose distribution is described in Table \ref{tab:1} \cite{servadio2020recursive,carravetta1997polynomial}. 

\begin{table}[h!]
\centering
\caption{Measurement noise distribution (rad)}
\begin{tabular}{c| c c c}
\toprule
$\eta_k$ & $1\mathrm{e}{-3}$ & $-3\mathrm{e}{-3}$ & $-9\mathrm{e}{-3}$ \\
\midrule
$P(\eta_k)$ & $\dfrac{15}{18}$ & $\dfrac{2}{18}$ & $\dfrac{1}{18}$ \\
\bottomrule
\end{tabular}
\label{tab:1}
\end{table}

\begin{figure}  [htbp!]
    \centering
    \includegraphics[width=.5\linewidth]{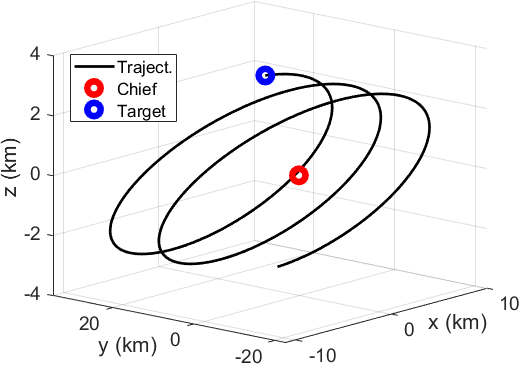}
    \caption{Target's relative motion with respect to the chief satellite.}
    \label{fig:traj}
\end{figure}

This application has been chosen since it presents a linear dynamic with nonlinear measurements. The linear system guarantees the exact uncertainty propagation of the state distribution, since a Gaussian PDF remains Gaussian after a linear transformation, and the standard Kalman filter represents the optimal solution if noises were Gaussians. Therefore, the difference and improvement in accuracy are due to the different order of measurement update. Figure \ref{fig:traj} shows the relative motion of the target with respect to the chief, creating a helical pattern around the origin whose amplitude is connected to the mean motion of the reference orbit. 

\begin{figure*}  [htbp!]
    \centering
    \includegraphics[width=0.93\linewidth]{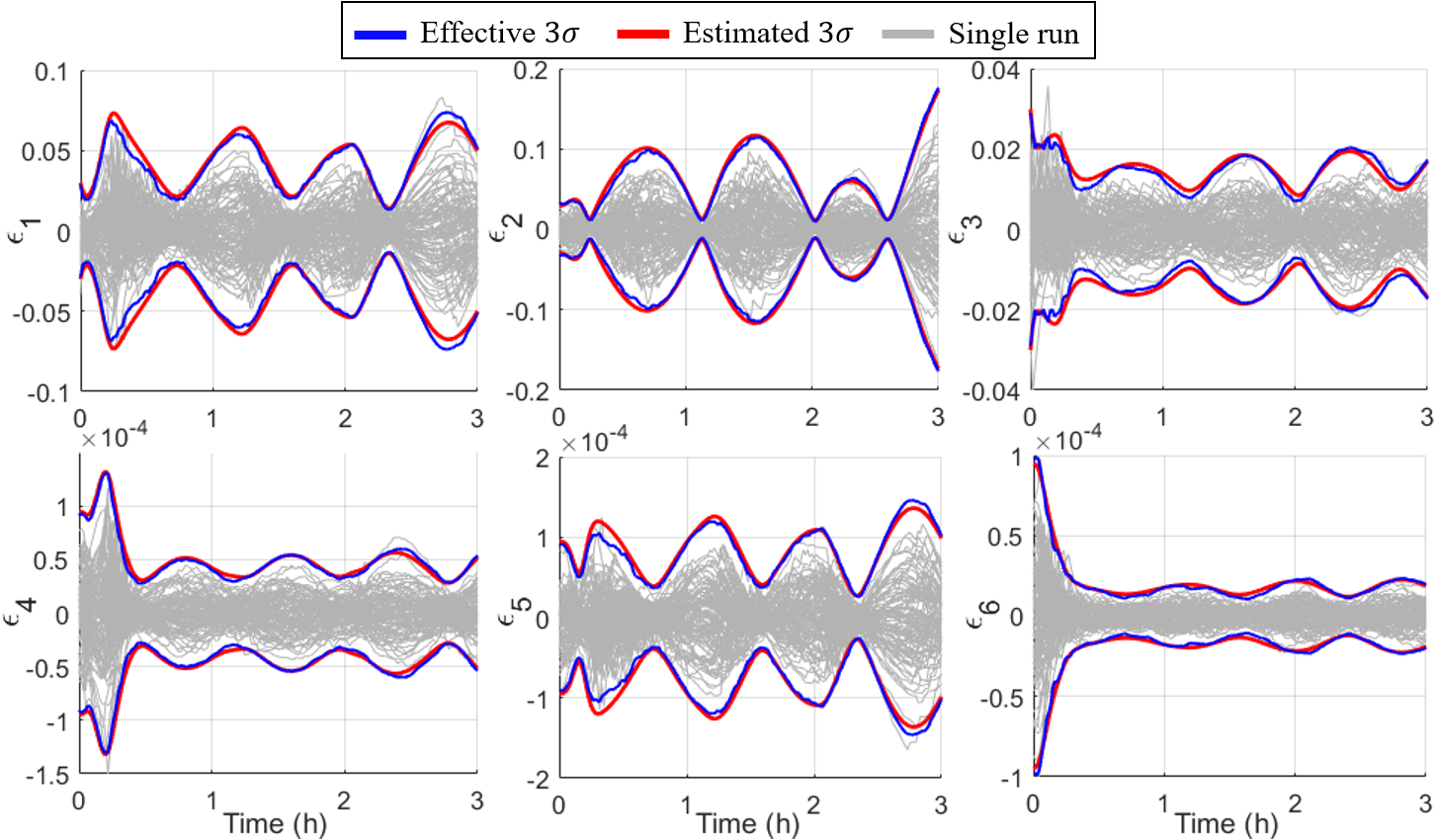}
    \caption{Monte Carlo Consistency Analysis for the Convergence of the QUKF.}
    \label{fig:mc}
\end{figure*}
The accuracy level and consistency of the filter are evaluated by Monte Carlo analysis, reported in Fig. \ref{fig:mc} for the QUKF. In the figure, the position and velocity state errors for each run are evaluated at each time step, as
\begin{equation}
    \epsilon_i = \hat{x}_i - x_{T}
\end{equation}
Each gray line is a separate simulation. The \textit{estimated} covariance, in terms of the $3\sigma$ boundary, is calculated directly from the diagonal entries of the updated covariance matrix of the filter, and is reported in red. These lines indicate the filter's own estimation of its error spread, and indicate how confident it is in its estimate. In comparison, the \textit{effective} covariance level, again in terms of the $3\sigma$ boundary, represents how the filter accuracy actually performs, as these levels of standard deviation are calculated directly from the Monte Carlo runs. A consistent filter has an overlap between the estimated and effective error covariance levels, meaning that it is able to correctly predict its own uncertainties in the estimates it provides. Thus, the QUKF is a consistent estimator that achieves the correct tracking of the target motion. QCUKF-4 and CCUKF-6 behave similarly to QUKF, and therefore, the Monte Carlo analysis has not been reported.

\begin{figure*}  [htbp!]
    \centering
    \includegraphics[width=.85\linewidth]{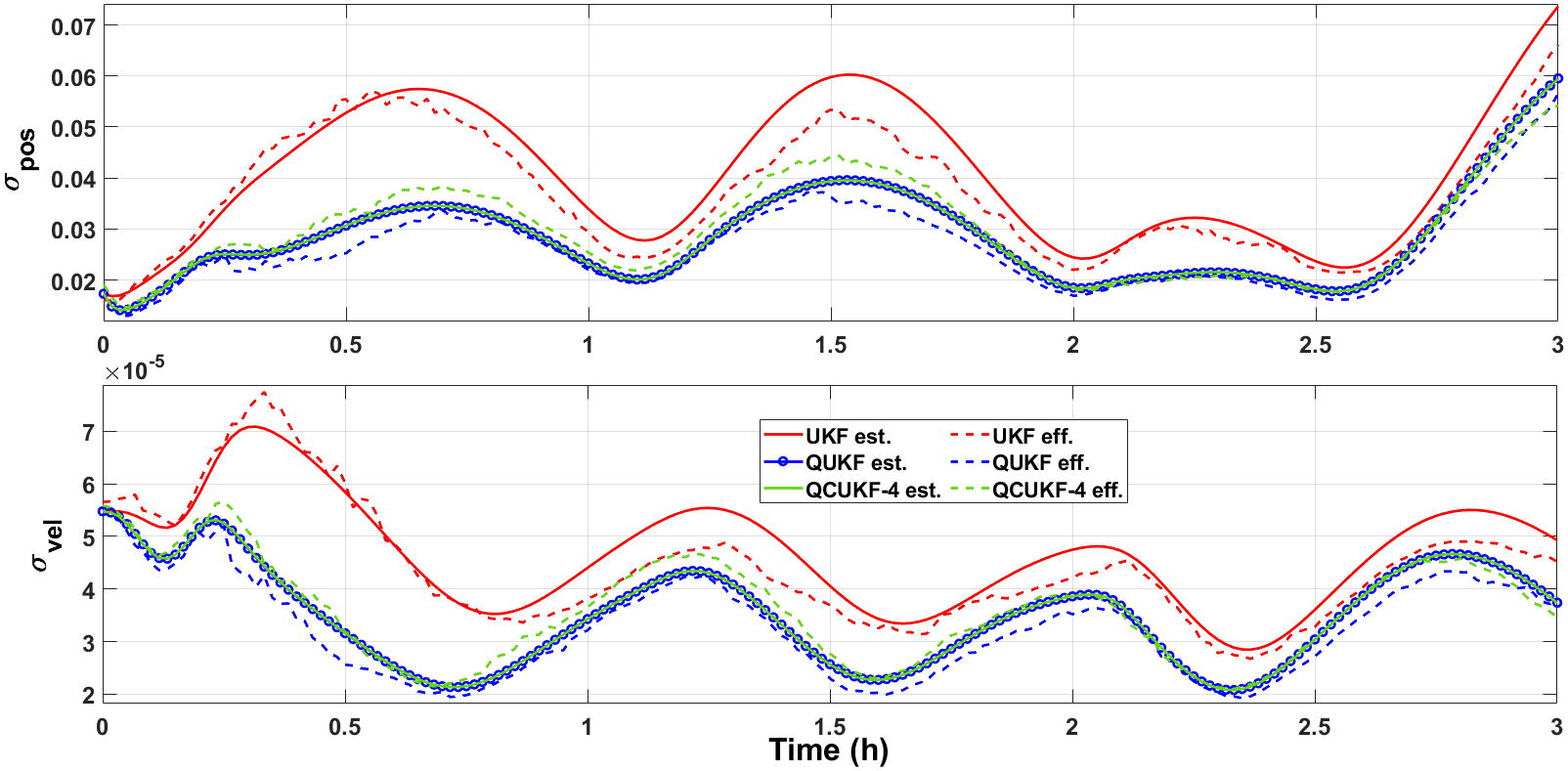}
    \caption{Error Standard Deviation Comparison between the linear estimators UKF and the quadratic estimators QUKF and QCUKF-4.}
    \label{fig:comp}
\end{figure*}

Now that the consistency of QUKF has been assessed, it is of interest to compare its precision to its linear counterparts. Figure \ref{fig:comp} shows the advantages in accuracy obtained by a QMMSE filter over an LMMSE one. The figure reports the error standard deviation levels both for position and velocity. Estimated covariances are evaluated directly from the updated step of the filter as 
\begin{align}
    \sigma_{pos, EST} &= \sqrt{P_{xx}+P_{yy}+P_{zz}} \\
    \sigma_{vel, EST} &= \sqrt{P_{v_xv_x}+P_{v_yv_y}+P_{v_zv_z}} 
\end{align}
while the effective error standard deviation, coming from the Monte Carlo analysis with multiple runs, is evaluated as
\begin{align}
    \sigma_{pos, EFF} &=  \sqrt{\sum_{j=\{x,y,z\}}\Bigg(    \sum_{i = 1}^{N_{MC}}(\epsilon_{j,i}- \hat{\epsilon_j})^2 \Bigg)   } \\
    \sigma_{vel, EFF} &=  \sqrt{\sum_{j=\{v_x,v_y,v_z\}}\Bigg(    \sum_{i = 1}^{N_{MC}}(\epsilon_{j,i}- \hat{\epsilon_j})^2 \Bigg)   }  
\end{align}
for each time step of the simulation. Once again, a consistent filter is assessed by the overlapping of the two standard deviations, reported as continuous and dashed lines in the figure, respectfully. At first sight, the QUKF (in blue) and the QCUKF-4 (in green) show evident advantages in accuracy, with error levels well below those of the UKF (in red). The two quadratic update estimators better account for the non-Gaussianity of the noise distribution. Thanks to the knowledge and inclusion of the non-Gaussian noise high order central moments, both the QUKF and the QCUKF-4 outperform the UKF, which has no information regarding the shape of the noise PDF and, therefore, assumes it as Gaussian.

%%%%%%%%%%%%%%%%%%%%%%%%%%%%%%%%%%%%%%%%%%%%%%%%%%%%%%%%%%%%%%%%%%%%%%
\subsection{Circular Restricted 3 Body Problem}
\begin{figure}  [htbp!]
    \centering
    \begin{subfigure}{0.48\textwidth}
        \centering
        \includegraphics[width=.45\textwidth]{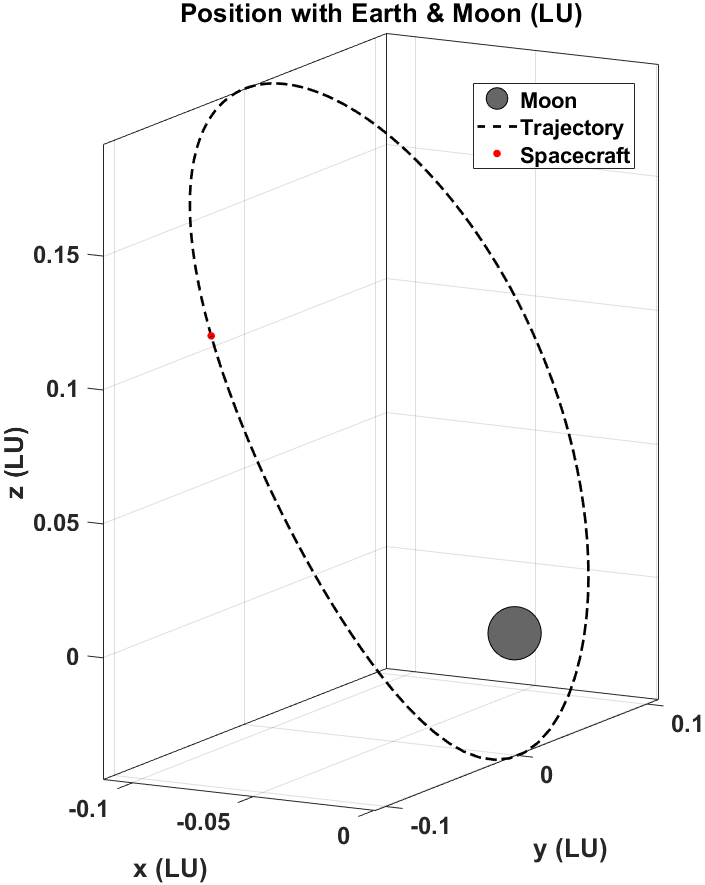}
        \caption{}
        \label{fig:cr3bp1}
    \end{subfigure}
    \hfill
    \begin{subfigure}{0.48\textwidth}
        \centering
        \includegraphics[width=.45\textwidth]{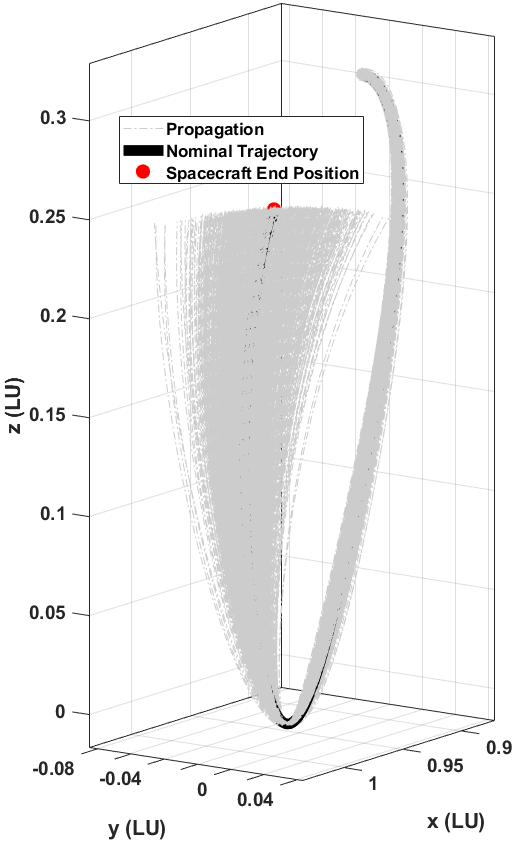}
        \caption{}
        \label{fig:cr3bp2}
    \end{subfigure}
    \caption{Example of a zoomed-in trajectory of a satellite in the CR3BP Earth-Moon system (a) and trajectory spread of CR3BP Monte Carlo simulation(b)}
    \label{fig:cr3bp}
\end{figure}

In order to observe the behavior of the higher order filters, we put them against a considerably harder nonlinear system to test their robustness. Simulations were performed in many dimensionless planetary systems. In this paper, we will focus on the Earth-Moon system. The synodic rotating frame is centered at the Earth-Moon barycenter, with the x-axis pointing from Earth to the Moon, the z-axis aligned with the angular momentum of the system, and the y-axis completing the right-handed triad. Distances are normalized by the Earth-Moon distance ($L$), and time is normalized by 
\begin{align}
    T^* &= \sqrt{\dfrac{(L^*)^3}{\mu_g(m_E + m_M)}},
\end{align}
so that the non-dimensional mean motion is unity. The corresponding length and time units are denoted by LU and TU respectively. The Earth-Moon mass parameter is taken as
\begin{align}
    \mu &= \dfrac{m_M}{m_E + m_M} = 1.215058560962404 \times 10^{-2},
\end{align}
so that in the rotating barycentric frame, the Earth and Moon are located at \((x,y,z) = (-\mu,0,0)\) and \((1-\mu,0,0)\) respectively. 
The system state is $\mbf x = [\mbf r \  \mbf v]^T =[ x \ y \ z \ \dot{x} \ \dot{y} \ \dot{z}]^T$, expressed in LU and LU/TU. The equations of motion follow the standard CR3BP form,
\begin{align}
    r_1 &= \sqrt{(x + \mu)^2 + y^2 + z^2},\\
    r_2 &= \sqrt{(x-1 + \mu)^2 + y^2 + z^2},\\
    U_x &= x - (1-\mu)\dfrac{x+\mu}{{r_1}^3} - \mu\dfrac{x-1+\mu}{{r_2}^3},\\
    U_y &= y - (1-\mu)\dfrac{y}{{r_1}^3} - \mu\dfrac{y}{{r_2}^3},\\
    U_z &= - (1-\mu)\dfrac{z}{{r_1}^3} - \mu\dfrac{z}{{r_2}^3}
\end{align}
with the state dynamics;
\begin{align}
    \dot{x} &= v_x,\\
    \dot{y} &= v_y,\\
    \dot{z} &= v_z,\\
    \dot{v}_x &= 2 v_y + U_x,\\
    \dot{v}_y &= -2 v_x + U_y,\\
    \dot{v}_z &= U_z
\end{align}
These equations are implemented directly in the numerical propagator as a first-order ODE system. We used a northern halo orbit trajectory about the $\mathbf{L}_1$ libration point in the Earth-Moon system, shown in Fig. \ref{fig:cr3bp1}. The initial condition and orbital period are taken from NASA's JPL periodic orbit catalog for the three-body problem, specifically the Earth-Moon, Halo, Northern, $\mathbf{L}_1$ family (orbit 896 in the online database\footnote{NASA JPL, "Periodic Orbits in the Circular Restricted Three-Body Problem"}). The non-dimensional orbital period for this orbit is $T_{per} = 2.1783120807931518$ TU, and the corresponding initial state is \begin{equation}
\mathbf{x}_0 =
    \begin{bmatrix}
    0.87592140310093525 \\ 
    -1.59031517986629 \times 10^{-26} \\
    0.19175810982939320 \\
    -2.93025310878967 \times 10^{-14} \\ 
    0.23080031482213192 \\
    7.3649704261223776 \times 10^{-14}
    \end{bmatrix},
\end{equation}
given in LU and LU/TU in the rotating barycentric frame. 

For the Monte Carlo experiments, the truth trajectory is generated by numerically integrating the CR3BP equations for the $N_{rev} = 2$ orbital periods, i.e, $t \in [0, \space t_f], \quad t_f = N_{rev}T_{per}$ with a sampling interval $ \Delta t = T_{per}/20.$ State propagation is performed using the variable-step \texttt{ode45} (Runge-Kutta-Fehlberg) MATLAB's integrator with relative and absolute tolerances of $10^{-12}$. Figure \ref{fig:cr3bp2} shows the possible 100 trajectories taken by the satellite. 

To emulate unmodeled accelerations and modeling errors, zero-mean Gaussian process noise is added to the truth dynamics. The continuous-time process noise covariance is chosen as $\mathbf{Q} = 10^{-7} \mathbf{I}_3$, so that the injected perturbations primarily affect the velocity components at each propagation step. In the Monte Carlo implementation, process noise is applied additively to the velocity states at each sampling time:
\begin{align}
    \mathbf{v}({t_k}^+) = \mathbf{v}({t_k}^-) + \mathbf{L}_q \mathbf{w}_k, \quad \mathbf{w}_k \sim \mathcal{N}(\mbf 0, \mathbf{I}_3),
\end{align}
where $\mathbf{L}_q$ is a Cholesky factor of the velocity block of $\mathbf{Q}$.

Measurements are constructed as line-of-sight range and range-rate evaluated from the system barycenter. The range and range rate measurement models are  
\begin{align}
    \mathbf{\rho} = \left| \mathbf{r}\right|, \quad \dot{\rho} = \dfrac{ \mathbf{r}^T  \mathbf{v}}{\mathbf{\rho}},
\end{align}
Thus, the measurement vector at time $t_k$ provided to the filter includes noise
\begin{align}
    \mathbf{y}_k = 
    \begin{bmatrix}
        \rho_k &
        \dot{\rho}_k
    \end{bmatrix}^T
    \space + \space \mathbf{v}_k, \quad \mathbf{v}_k \sim \mathcal{N}(\mathbf{0}, \mathbf{R}),
\end{align}
with a diagonal measurement covariance $\mathbf{R} = \mathrm{diag}({\sigma_r}^2, \sigma_{\dot{r}}^2)$, with $\sigma_r = 10^{-3} \mathrm{LU}$ and $\sigma_{\dot{r}} = 10^{-3} \mathrm{LU/TU}$. Higher-order central moments of the measurement noise (third- and fourth order) are also computed using the Isserlis theorem to support the polynomial update filter. The initial covariance of the state is taken as $ \mathbf{P}_0 = \mathrm{blkdiag}(10^{-5}\mathbf{I_3},10^{-5}\mathbf{I_3})$, which corresponds to small uncertainties in both position and velocity about the nominal halo orbit. 

\begin{figure*}  [htbp!]
    \centering
    \includegraphics[width=.93\linewidth]{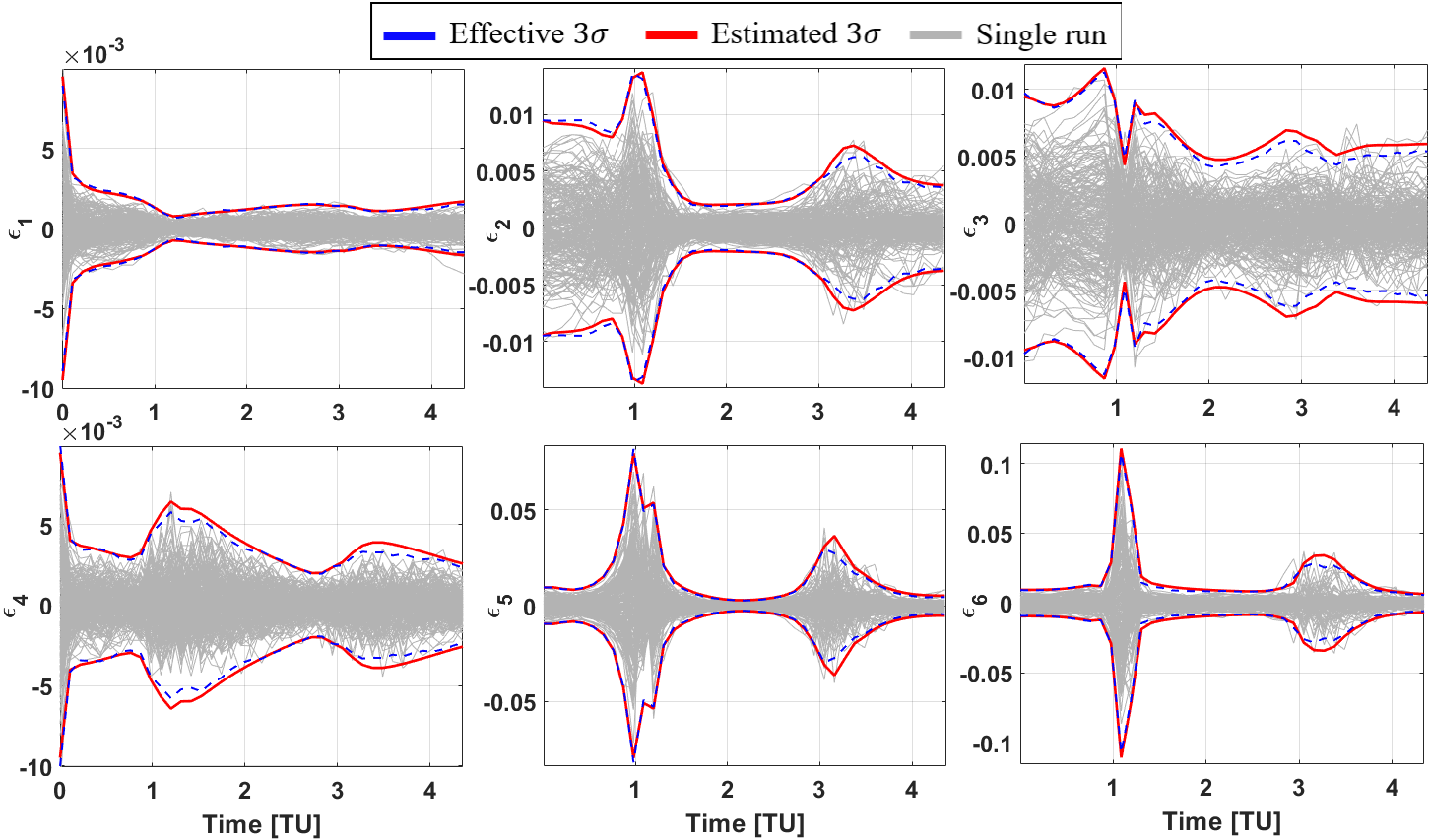}
    \caption{Monte Carlo Consistency Analysis for the Convergence of the CACUKF-6}
    \label{fig:cr3bp_cacukf}
\end{figure*}

Similarly to the Clohessy-Wiltshire application, consistency checks were implemented via Monte Carlo analysis. The position and velocity state errors were calculated in the same way as in the previous case, as shown in Fig. \ref{fig:cr3bp_cacukf}. The figure shows how the CACUKF-6 filter slowly converges the uncertainties even in highly unstable orbits or with high levels of Gaussian noise. The lower-order filters, such as the UKF, do not converge as well as the QACUKF-4 and CACUKF-6 filters. 

\begin{figure*}  [htbp!]
    \centering
    \includegraphics[width=.93\linewidth]{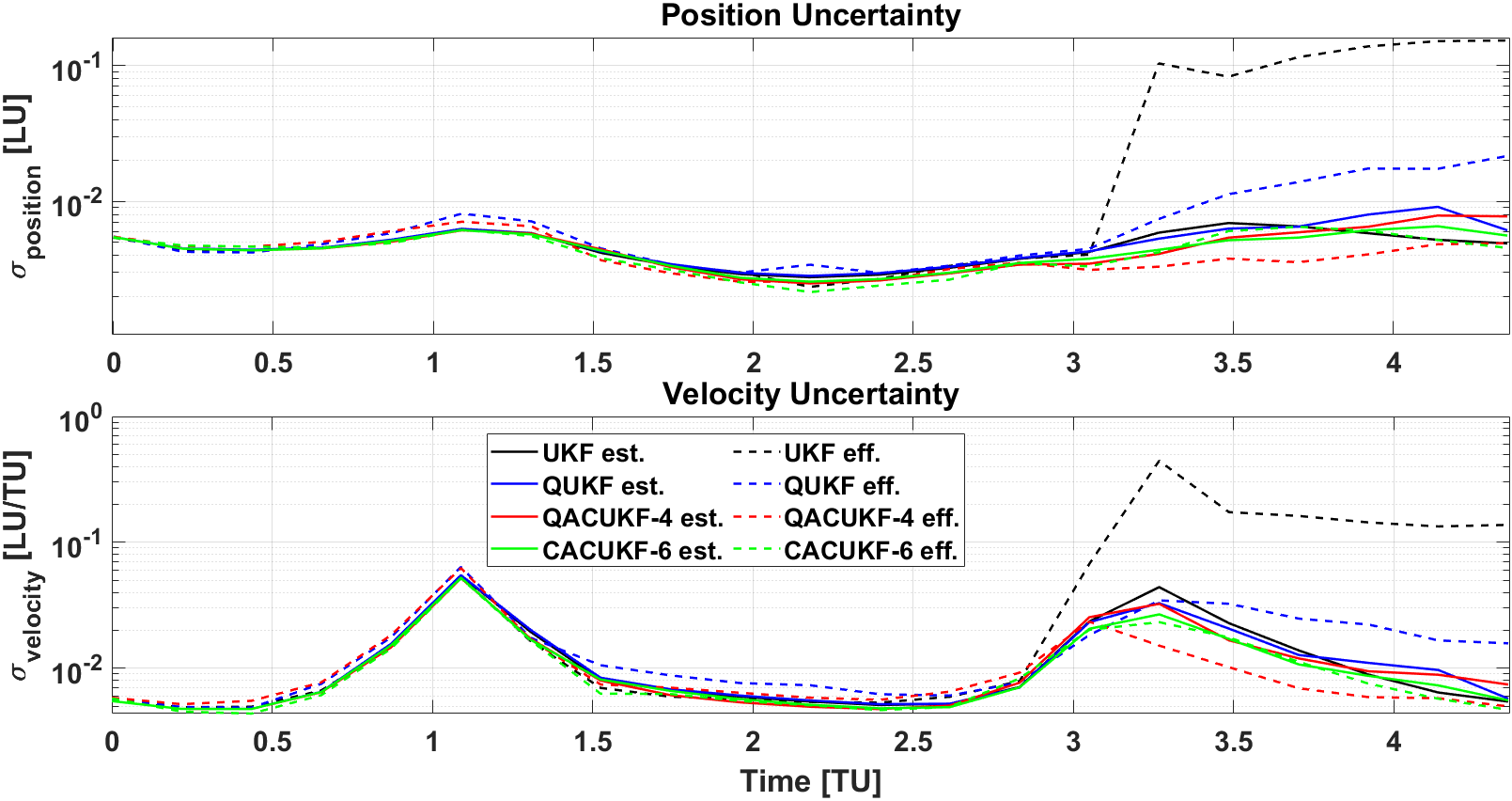}
    \caption{Error Standard Deviation Comparison between the linear estimator (UKF), the Quadratic Estimators (QUKF and QACUKF-4) and the Cubic Estimator (CACUKF-6).}
    \label{fig:cr3bp_error}
\end{figure*}

To observe the quantitative difference among the UKF, QUKF, QACUKF-4, and CACUKF-6, Fig. \ref{fig:cr3bp_error} compares their respective Monte Carlo analysis. This figure compares the precision of the filters similar to the error standard deviation comparison for the CW case. The CACUKF-6 (green lines) shows the highest consistency and robustness in precision, followed by QACUKF-4 (red lines), as highlighted in the last section of the simulation when the trajectory of the spacecraft considerably leaves the periodic halo orbit. The UKF (black lines) loses tracking of the spacecraft, diverging right before 3 TU. Lastly, the QUKF (blue lines) is more accurate than the UKF, but not quite as precise as the filters embedded with the CUT.

The filters proposed in this paper provide a natural improvement of the UKF in the sense that a parabolic fitting provides a more accurate representation than the linear approximation. We were able to obtain a quadratic approximation of the MMSE using well-known uncertainty propagation techniques such that the resulting QUKF and QAUKF are easily accessible to the general reader as new filter benchmarks. Moreover, the quadratic formulation of the update, which is the core of the QMMSE, can be expanded to any uncertainty transformation technique, as the Kronecker formulations provided work regardless of how $\mbf P_{yy}$ has been evaluated. Replacement of the unscented transformation with CUT improves the robustness and accuracy of the polynomial filters. An increase in the number of sigma points allows these filters to excel in highly nonlinear environments, provided that the calculations of radii and weights are done correctly depending on the dimension of the predicted and measured states.

\section{Conclusion}
The paper derived a quadratic approximation of the true MMSE, implemented alongside the standard UKF mathematics. That is, the proposed QUKF is the parabolic improvement of the linear UKF, so that the estimator better follows any curved shape of the true posterior distribution. The theoretical derivation of the quadratic estimator, approximation of the QMMSE, is extendable to any moment propagation technique, as long as they accurately provide a reliable evaluation of expectations. Indeed, the QUKF and QAUKF have been improved by including the conjugate unscented transformation to replace the standard UT, yielding more robust and consistent filters: QCUKF-$k$ and QACUKF-$k$. 

The quadratic update approximates the true MMSE more accurately than the linear update, especially
when influenced by non-Gaussian noise that requires information of high-order central moments, such as skewness and kurtosis. The QUKF and QACUKF-4 obtain additional information during the evaluation of the augmented Kalman gain, thereby providing a more accurate estimate, as they better understand the influence of the measurement noise distribution.

Conceptually, the technique can be extended to any arbitrary high-order polynomial approximation of the update, as shown by the cubic representation of the MMSE: the CACUKF-$k$ filter, which is based on the CMMSE derivation. It requires a precise, attentive evaluation of the noise's influence on expected-value estimates whenever dealing with the residual. In fact, as the order of the update increases, the knowledge of higher-order central moments of the distributions must be provided or approximated to guarantee the correct evaluation of the augmented Kalman gain and the proper functionality of the filter. 

Regarding the benefits in estimation shown by the proposed numerical applications, the QUKF and QAUKF achieve significant improvements using well-known techniques and minor changes to the original UKF algorithm, so that the gains in accuracy come at a limited additional computational cost. When required by a more complex scenario with particularly strong nonlinearities, the inclusion of the CUT (QACUKF-$k$ and CACUKF-$k$) ensures higher accuracy levels.

\bibliography{references}

\end{document}